\documentclass[
reprint,
 amsmath,amssymb,
 aps,
floatfix,
]{revtex4-2}
\DeclareMathOperator{\sinc}{sinc}
\usepackage{hyperref}
\hypersetup{colorlinks,linkcolor={blue},citecolor={blue},urlcolor={red}} 
\usepackage{physics}
\usepackage{mathtools}  
\DeclarePairedDelimiter{\ceil}{\lceil}{\rceil}
\usepackage{ulem}
\usepackage[dvipsnames]{xcolor}

\usepackage[x11names]{xcolor}
\usepackage{subcaption}
\usepackage{graphicx}

\usepackage{float}
\usepackage[dvipsnames,table]{xcolor}  
\definecolor{mycolor}{rgb}{0.5,0.,0.5}
\definecolor{violet}{rgb}{0.6,0,0.6}
\newcommand{\DB}{\textcolor{black}}

\usepackage{graphicx}
\usepackage{dcolumn}
\usepackage{bm}

\begin{document}
\preprint{APS/123-QED}
\title{On the influence of optical smoothing techniques on cross-beam energy transfer }

\author{Y. Lalaire$^{1-2}$}\email{yann.lalaire@cea.fr}
\author{C. Ruyer$^{1-2}$} \email{charles.ruyer@cea.fr}
\author{A. Debayle$^{2-3}$} 
\author{G. Bouchard$^{1-2}$} 
\author{A. Fusaro$^{1-2}$} 
\author{P. Loiseau$^{1-2}$} 
\author{L. Masse$^{1-2}$} 
\author{P. E. Masson-Laborde$^{1-2}$} 
\author{D. Bénisti$^{1-2}$}
\affiliation{$^1$CEA, DAM, DIF, F-91297 Arpajon, France}
\affiliation{$^2$Université Paris-Saclay, CEA, LMCE, 91680 Bruyère-Le-Chatel, France}
\affiliation{$^3$Focused Energy GmbH, Im Tiefen See 45, 64293 Darmstadt, Germany}
\date{\today}

\begin{abstract}
In the context of inertial confinement fusion (ICF) experiments, spatial and temporal laser beam smoothing techniques are used to control the beams propagation in hohlraum plasmas.
Currently, spatial and temporal smoothing are either neglected or not properly taken into account in the inline cross beam energy transfer (CBET) models included in the hydrodynamic codes dedicated to the design of these experiments. In some cases, which we will highlight in this study, this simplification leads to important errors in the power transfer of importance for the implosion symmetry of the capsule,  either in the direct or indirect drive ICF configurations.  
In a recent study [A. Oudin et \textit{al}., Phys. Plasmas \textbf{32}, 042706 (2025)], we demonstrated the necessity of accounting for spatial smoothing when modeling CBET, provided that the beams do not have the same wavelength.
This work presents a linear kinetic model compared with Hera paraxial fluid simulations and compared with the Smilei particle-in-cell code, demonstrating the important influence of smoothing by spectral dispersion on CBET. Moreover, we demonstrate the importance of accounting for the plasma velocity profile, the beam modulation bandwidth, and the spectral dispersion to better predict the power exchanged between the beams. Additionally, we reveal a strong sensitivity of this power transfer to the synchronization of the phase modulators. 
\end{abstract}
\maketitle
\captionsetup{labelformat=empty}
\section{\label{I Introduction} Introduction}
Smoothing by spectral dispersion~\cite{Skupsky,Rose1,Kato} introduces spatial and temporal variations in the field phase to achieve improved control over laser beam propagation~\cite{Moody} in inertial confinement fusion (ICF) experiments~\cite{Abu-Shawa}. 
It has been shown that optical smoothing of laser beams modifies the growth rate of parametric instabilities~\cite{Berger1,RuyerSBS,RuyerSpray,Moody} (stimulated Raman/Brillouin backscatter), which are known to cause laser beam energy losses~\cite{Zylstra1,Zylstra2}. Particularly, Cross-Beam Energy Transfer (CBET) can be affected by these optical smoothing techniques. In indirect drive, CBET occurs at the entrance of the hohlraum in a highly damped plastic plasma, resulting from the ablation of the hohlraum entrance window, or closer to the inner walls of the cavity in a very weakly damped gold plasma. Controlling this exchange of energy between beams is crucial, as it ensures uniform compression of the fuel capsule held in the centre of the hohlraum~\cite{Kritcher_PRE,Abu-Shawa,Kritcher_NATURE,Hao_2025}. Likewise, CBET also affects the dynamics of a  directly driven capsule \cite{Igumenshchev_2010}. The CBET drives low mode perturbations on the implosion sphericity modifying neutron yields and bang times \cite{PRL_Colaitis_2022}.\\
Most high energy laser facilities use spatial smoothing by phase plates (RPP) which consists in decomposing the laser beam into small speckles. Some facilities also use polarization smoothing (PS) which decomposes the light onto two decorrelated speckle patterns.
At the National Ignition Facility (NIF), the Omega facility (LLE), and the Gekko XII facility (ILE), Transverse Smoothing by Spectral Dispersion (TSSD) is further applied to effectively control laser-plasma instabilities (LPI)~\cite{Skupsky}. At the Laser Megajoule, Longitudinal Smoothing by Spectral Dispersion (LSSD) is used to mitigate these scatterings.  Smoothing by spectral dispersion induces a periodic displacement of the speckles by modulating the laser beam frequency, thereby broadening the temporal spectrum of the laser beam. The different frequency components, or colors, are then spatially dispersed across the phase plate. The distinction between transverse smoothing by spectral dispersion and longitudinal smoothing by spectral dispersion lies in the direction in which the dispersion of these frequency components is applied~\cite{Videau_JOSA,Garnier_Videau_2001}.\\ 
Hydrodynamic codes dedicated to the design of ICF experiments mostly incorporate inline CBET models that consider the crossing of two plane waves~\cite{Strozzi,Debayle_rt,Liberatore_2023}. 
In Ref.~\cite{OudinPOP2}, a 2D model of spatially smoothed beams is introduced, thereby neglecting 3D effects as well as the phase modulation and dispersion induced by SSD. In Ref.~\cite{Michel,Michel_POP_2009}, a model with spatially and temporally smoothed beams is employed, but phase dispersion is neglected. In Refs.~\cite{Seaton1, Seaton2}, a model is presented where both the aperture imposed on the beams by the focusing optics and the phase dispersion are neglected. \\ 
In this work, we consider the crossing of two spatially and temporally smoothed beams in 3D, approaching a realistic configuration for ICF experiments. We propose a model, supported by both particle-in-cell (PIC) and paraxial hydrodynamic simulations, that simultaneously accounts for beam aperture, phase modulation, and phase dispersion, and aiming at being integrated into simulation codes. We show that our model predicts a power exchange significantly different from usual plane-wave models. In particular, we highlight the strong dependence of CBET on the laser chain synchronization and the phase dispersion. We demonstrate the need to account for the effect of the laser smoothing in 3D, especially for weakly damped plasmas—an aspect not addressed in the existing literature to the best of our knowledge.
We also demonstrate that the power exchange predicted in Ref.~\cite{Michel} corresponds to an average over the modulator synchronization, and that a lack of control of this quantity introduces inaccuracies of the CBET predictions. The companion paper Ref. \cite{Lettre_lalaire} simplifies the power transfer formulas to extract a simple criterion  indicating the laser and plasma parameters for which the CBET predictions deviate from a plane wave model. \\
The present paper is organized as follows. In Sec.~\ref{II Heuristique Interprétation}, we first discuss how power transfer between speckled beams differs from the plane-wave case. The comparison between the ponderomotive grating resulting from the crossing of the two laser beams and the acoustic wave profile will pinpoint how optical smoothing affects CBET. 
Section~\ref{III Modele} describes the modeling of the crossing of two spectrally smoothed beams within a plasma. We describe the ponderomotive excitation and the acoustic response of the plasma using the linear kinetic theory applied to these beams. This leads to a fairly simple expression for the exchanged power when the time derivative can be neglected in the electromagnetic wave equations. We examine the influence of SSD on CBET, with particular emphasis on the necessity of accounting for spectral dispersion induced by a dispersive grating or a focusing grating, as well as the synchronization of SSD between the two laser chains. 
Section~\ref{V Comparaison Hydro} compares our model predictions against results obtained using the paraxial--hydrodynamic code \textsc{HERA}~\cite{Loiseau_2006,Hera1,Hera2}, while Sec.~\ref{VI Comparaison PIC} compares our predictions against those from the PIC code \textsc{Smilei}~\cite{Smilei}.
The final section, Sec.~\ref{VI Conclusion et perspectives}, is dedicated to our conclusions.

\section{\label{II Heuristique Interprétation}Influence of optical smoothing on CBET}
This section aims to give a qualitative understanding of the influence of optical smoothing on CBET. 
To this end, we use the model presented in Sec.~\ref{III Modele} and compared  with paraxial--hydrodynamic and PIC simulations in Sec.~\ref{V Comparaison Hydro} and Sec.~\ref{VI Comparaison PIC}, respectively.
The spatial and temporal smoothing consists in degrading the spatial and temporal coherence of the laser beam. Hence, the spatially smoothed beam is composed of many slightly misaligned electromagnetic wavelets. When adding SSD, the beam is also the superposition of many slightly frequency-shifted wavelets. 
The phase of each of these wavelets is tilted by a grating in the case of the TSSD or has a different curvature when using a focusing grating in the case of the LSSD.
This contrasts with the so-called plane wave and monochromatic limit in which most CBET models are derived and where the laser beam is composed of a single wavevector and frequency component. With that in mind, the crossing of two smoothed laser beams results in the superposition of many gratings coupling each wavelet of the two laser beams and with many different beating wavectors and frequencies. \\
In this section, \DB{we discuss the stretching of the acoustic grating relative to the ponderomotive one (as done at acoustic resonance in Ref.~\cite{OudinPOP2}) to gain insight into the mechanisms that enhance or hinder the power exchange compared with the classical plane-wave crossing configuration.} \DB{We first address the single speckle configuration in Subsec.~\ref{II.1 Monospeckle}, then  RPP beams in Subsec.~\ref{II.2 Spatial smoothing} and, subsequently, realistic SSD beams in Subsec.~\ref{II.3 Temporal smoothing}. The results shown in this Section, and plotted in Figs.~\ref{Delocalisation}-\ref{TSSD pondero acoustique vdy}, are derived from  the fluid limit of the CBET plasma response detailed in Subsec.~\ref{III.3 Battement acoustique} and assume SSD-synchronized beams ($t_0=0$, see Part.~\ref{III.6 Influence synchronisation}). }

\subsection{\label{II.1 Monospeckle}Probing Acoustic Resonance within the Monospeckle Model}
\begin{figure}
\centering
\includegraphics[scale=0.30]{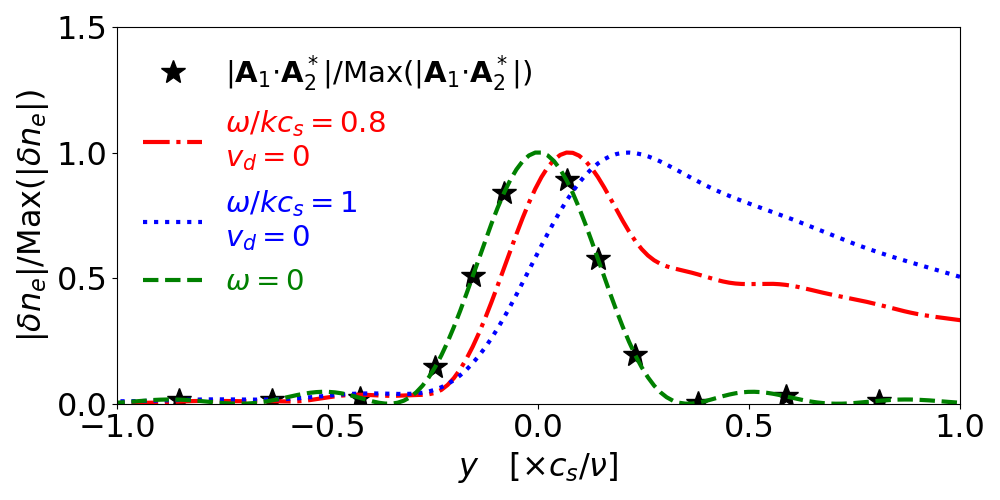}
\vspace{-0.6em}
\caption{\colorbox{Gray!20}{\parbox{\dimexpr\linewidth-2\fboxsep}{\textbf{FIG \thefigure:} Envelope of the ponderomotive (Eq.~\eqref{pondero beat}) and acoustic gratings (Eq.~\eqref{dne}) along the acoustic-wave propagation direction $(Oy)$ in a monospeckle configuration ($\phi_{\mathbf{n}_1} \equiv 0$ and $\phi_{\mathbf{n}_2} \equiv 0$), normalized to their maximum values. The case in which the beams have the same wavelength and intersect in a plasma with any flow along the $(Oy)$ direction is represented by the green dashed line. Situations where the plasma is stationary and \DB{when there is a frequency detuning,  $\omega/kc_s = 0.8, 1.0$, between the beams}  are represented by the red dash-dotted curves and blue dotted curves, respectively. We consider a carbon plasma and use: $N = 30$, $\theta = 12^\circ$, $\lambda_0 = 0.351\,\mu\text{m}$, $\textbf{u}_1 \parallel \textbf{u}_2$ and $f_\sharp=8$ (see Sec. \ref{III Modele}).}}}
\label{Delocalisation}
\end{figure}
\DB{Fig.~\ref{Delocalisation} plots, in the monospeckle configuration, the envelopes of the ponderomotive and acoustic gratings,  normalized to their maximum values, against the main direction of propagation of the acoustic wave, $y$. These envelopes are derived from Eq.~\eqref{pondero beat} and Eq.~\eqref{dne} with $\phi_{\mathbf{n}_1} = 0$ and $\phi_{\mathbf{n}_2} = 0$, respectively for the ponderomotive and acoustic gratings.}
We recover the result from Ref.~\cite{OudinPOP2}: \DB{a wavelength detuning between the beams}, near acoustic resonance, leads to a stretching of the acoustic grating envelope (blue dotted curve) relative to the ponderomotive grating envelope (black star markers). \\

The decoupling between the acoustic and ponderomotive gratings reduces the power exchanged between the smoothed beams compared to the plane-wave crossing model. If the IAW damping rate is large enough, either because of the plasma parameters or because the crossing is operated with a large enough angle, the IAW does not leave the speckle's waist and the power exchange should match the plane wave predictions. 
In contrast, \DB{when the beams have the same wavelengths}  (green dashed line), no stretching is observed, regardless of the plasma flow value along the $(Oy)$ direction.\\
The red dash-dotted curve in Fig.~\ref{Delocalisation} shows that this expansion occurs even out of resonance, and its extent decreases away from acoustic resonance. 
The ability of the acoustic wave to leave the vicinity of the crossing region when the beams are frequency-shifted explains why the resulting CBET differs from the plane wave limit, as explained in Ref. ~\cite{OudinPOP2}. \\
\DB{For beams with the same wavelength,} the absence of acoustic envelope expansion explains why the power exchange coincides with the  plane-wave predictions. The IAWs, propagating in the same direction as the plasma drift but in the opposite sense, remain stationary with respect to the ponderomotive grating envelope, similarly to the plane-wave crossing scenario.
When acoustic damping is reduced, the stretching increases because the IAWs are less damped during propagation. This explains why detuning beams has a dominant effect on weakly damped plasmas. \\
In the following, we will focus our study on acoustic resonance,  keeping in mind that the stretching of the acoustic grating envelope always represents a deviation in the exchanged power compared to the classic plane-wave scenario. Note that it results in a reduction in the exchanged power at resonance and an enhancement away from it.

\subsection{\label{II.2 Spatial smoothing}Influence of spatial smoothing on CBET}
\begin{figure}
\centering
\begin{subfigure}{0.23\textwidth}
    \centering
    \caption{$\forall(\mathbf{v}_d,\omega)$}\vspace{-0.4em}
    \includegraphics[width=\linewidth]{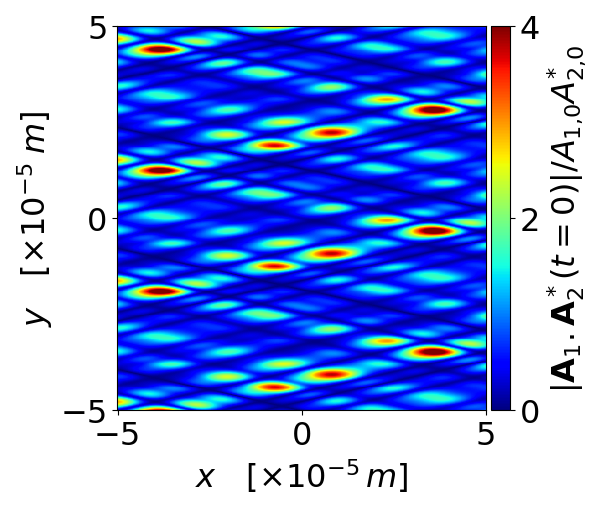}
    \label{RP RPP}
\end{subfigure}\hspace{0.15em}%
\begin{subfigure}{0.23\textwidth}
    \centering
    \caption{$\mathbf{v}_d=c_s\mathbf{e}_y$}\vspace{-0.4em}
    \includegraphics[width=\linewidth]{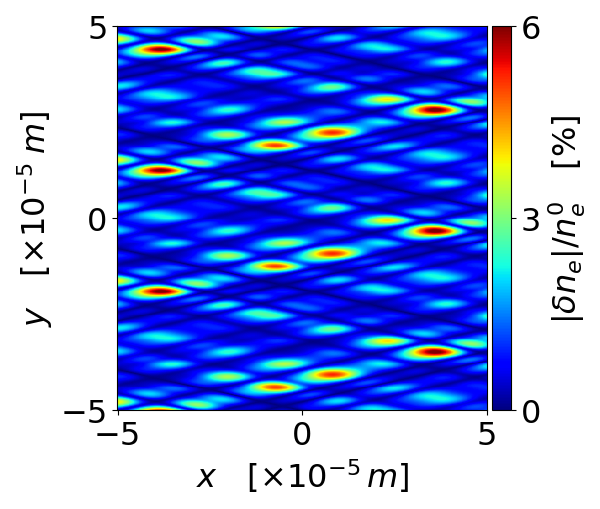}
    \label{RA RPP vdy}
\end{subfigure}\\[-0.5em] 
\begin{subfigure}{0.23\textwidth}
    \centering
    \caption{$\mathbf{v}_d/c_s=\mathbf{e}_y+3\mathbf{e}_x$}\vspace{-0.4em}
    \includegraphics[width=\linewidth]{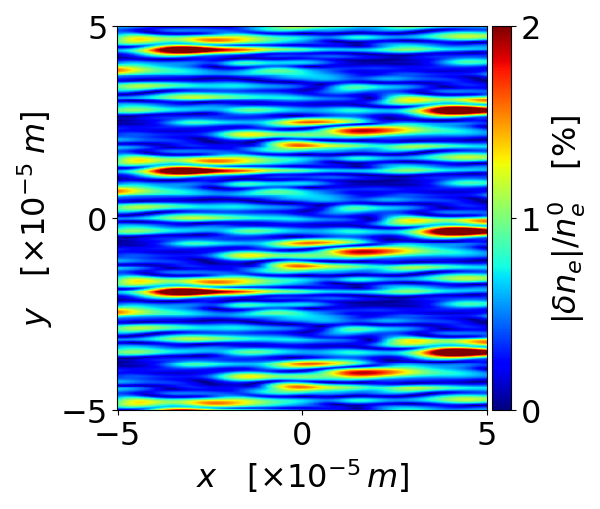}
    \label{RA RPP vdx}
\end{subfigure}\hspace{0.15em}%
\begin{subfigure}{0.23\textwidth}
    \centering
    \caption{$\omega_1-\omega_2=kc_s$}\vspace{-0.4em}
    \includegraphics[width=\linewidth]{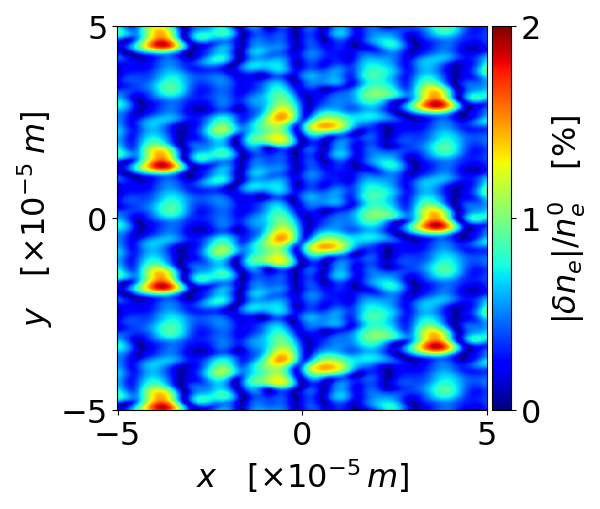}
    \label{RA RPP w}
\end{subfigure}
\vspace{-1.0em}
\caption{\colorbox{Gray!20}{\parbox{\dimexpr\linewidth-2\fboxsep}{\textbf{FIG \thefigure:} Maps in the $(xOy)$ plane of the RPP ($M=0$) ponderomotive grating envelope (a), the RPP acoustic grating envelope for two beams \DB{with the same wavelength} crossing in a carbon plasma drifting at the sound speed along $(Oy)$ (b), \DB{plasma drifting in the $(xOy)$ plane (c)}, and for the case where a wavelength detuning is imposed between the RPP beams (d). The only parameter that differs from Fig.~\ref{Delocalisation} is \( N = 10 \), and the laser vector potential \( A_{0,i} = 9.32 \times 10^{-6} \, \text{T} \cdot \text{m} \) (see Sec.~\ref{III Modele}).}}}
\label{RPP pondero acoustique vdy}
\end{figure}

We now consider the intersection of two spatially smoothed beams without SSD, produced by a one-dimensional phase plate composed of $N$ elements. The reduced number of phase plate elements ($N=10$) introduces an artificial periodicity in the interference pattern, as shown in Figs.~\ref{RPP pondero acoustique vdy} and \ref{TSSD pondero acoustique vdy}. This assumption has no impact on the subsequent qualitative analysis as the characteristic length of the damped IAW, $c_s/\nu$, is much smaller than the periodicity, $Nf_{\#}\lambda_0$.
Both laser beams are here composed of many wavelets with slightly misaligned wavevectors but equal frequencies. Figure ~\ref{RP RPP} illustrates the envelope of the ponderomotive grating and shows large amplitude regions corresponding to the crossing of the speckles that compose each beam.  When the two laser beams cross in a plasma, the ponderomotive grating results in a driven acoustic wave, which is illustrated in Fig.~\ref{RA RPP vdy} in the case of a $y$-aligned flow velocity. We demonstrate that the ponderomotive (Eq.~\eqref{pondero beat}, Fig.~\ref{RP RPP}) and acoustic (Eq.~\eqref{dne}, Fig.~\ref{RA RPP vdy}) gratings exhibit identical shape, with coinciding maxima. In this case, the driven acoustic wave does not leave the high field vicinity and its amplification is maximum \cite{OudinPOP2}, similarly to the crossing of two plane waves. In the Fourier space, all grating components which results from the coupling of all the two laser beam wavelets have no frequency (because the two laser beams have no frequency shift) so they are all resonant with the plasma drifting at the sound speed. Similarly to the crossing of two plane waves, here all the ponderomotive gratings are at resonance so that the power exchange is maximum. \\ 
Figures~\ref{RPP pondero acoustique vdy}(c,d) illustrate two scenarios where spatial smoothing disturbs CBET compared to the previous situation. The case where the plasma drifts with $\textbf{v}_d/c_s=\textbf{e}_y+3\textbf{e}_x$ while the beams have identical wavelength is illustrated in Fig.~\ref{RA RPP vdx}. Figure~\ref{RA RPP w} shows the envelope of the acoustic density fluctuation in the $(xOy)$ plane, when the plasma is stationary in the laboratory but the beams are wavelength-shifted. 
In both cases, we observe a stretching of the acoustic grating compared to the ponderomotive grating, a result of the propagation of the driven IAWs which allows only part of the grating components to be at resonance.
When the drift is perturbed in a direction other than $(Oy)$, for example, along the $(Ox)$ axis, the fluctuation is stretched in the $(Ox)$ direction, as shown in Fig.~\ref{RA RPP vdx}. 
Conversely, when a wavelength detuning is applied between the beams, we observe the stretching of the acoustic fluctuations in the $(Oy)$ direction, as shown in Fig.~\ref{RA RPP w}. 
In both cases, the amplification of the driven acoustic wave stops when it leaves the high field regions (see Fig.~\ref{RP RPP}), explaining the reduction in exchanged power at resonance for detuned RPP beams or for a plasma drifting in a different direction from that of the IAWs. 
Note that a similar mechanism is expected when $v_{dz}$ becomes comparable or larger than the sound speed.   
In essence, this section shows that accounting for spatial smoothing in CBET modeling becomes necessary only when the beams are wavelength-detuned or when they interact in a plasma exhibiting a flow component orthogonal to the IAW propagation direction.
\subsection{\label{II.3 Temporal smoothing}Influence of SSD temporal smoothing on CBET}
\begin{figure}
\centering
    \begin{subfigure}{0.24\textwidth}
        \centering
        \caption{$\forall(\mathbf{v}_d,\omega)$}\vspace{-0.4em}
        \includegraphics[width=\linewidth]{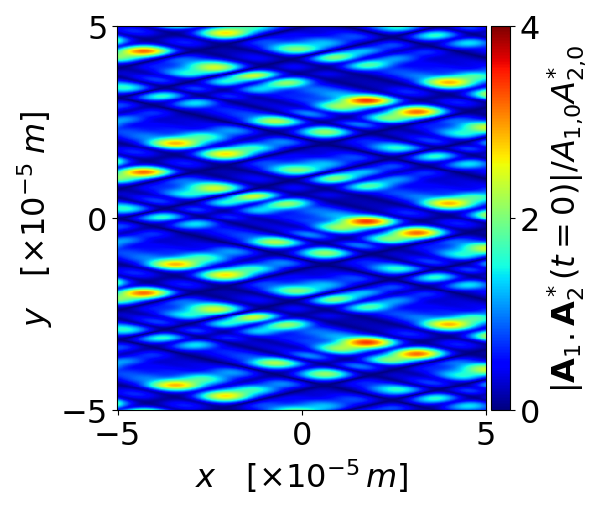}
        \label{RP TSSD}
    \end{subfigure}%
    \hfill
    \begin{subfigure}{0.24\textwidth}
        \centering
        \caption{$\mathbf{v}_d=c_s\mathbf{e}_y$}\vspace{-0.4em}
        \includegraphics[width=\linewidth]{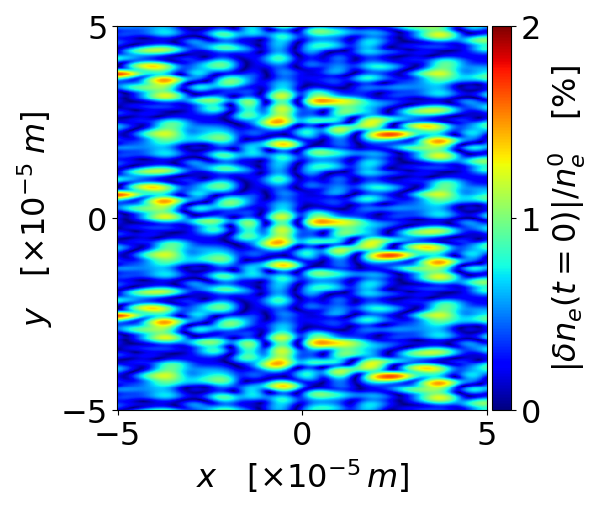}
        \label{RA TSSD}
    \end{subfigure}
    \vspace{-3.0em}
\caption{\colorbox{Gray!20}{\parbox{\dimexpr\linewidth-2\fboxsep}{\textbf{FIG \thefigure:} Maps in the $(xOy)$ plane of the TSSD ponderomotive beat envelope (a), and of the TSSD acoustic grating envelope (b) for beams \DB{with the same central frequency} crossing in a plasma drifing at the sound speed in the main IAW direction $(Oy)$. The only parameters that differ from Fig.~\ref{RPP pondero acoustique vdy} are: $M = 3.9$, $\Delta=\ceil{M}$, $\nu_d=17\,\text{GHz}$, $N_c=1$ and $t_0=0$ (see Sec. \ref{III.1 Champs rpp ssd}).}}}
\label{TSSD pondero acoustique vdy}
\end{figure}
We now illustrate the case of beam crossing when SSD is applied. We plot in the $(xOy)$ plane the envelope of the ponderomotive grating in Fig.~\ref{RP TSSD} and of the acoustic grating in Fig.~\ref{RA TSSD} in a plasma drifting at $c_s$ along the $(Oy)$ direction. We observe that these two gratings differ from the case without SSD shown in Fig.~\ref{RPP pondero acoustique vdy}. In contrast to the RPP case, we observe a stretching of the maxima of the acoustic grating relative to the ponderomotive grating. This stretching reflects the ability of the acoustic grating to preserve the memory of the speckle motion and explains the reduced power exchange near the acoustic resonance and its enhancement away from it for TSSD/LSSD beams, compared to the plane-wave case [see Figs.~\ref{Smilei comp}(a,b)]. For beams that are both spatially and temporally smoothed, the stretching induced by dispersion illustrated in Fig.~\ref{RA TSSD} competes with the stretching induced by wavelength detuning or by drift orthogonal to the IAWs shown in Fig.~\ref{RPP pondero acoustique vdy}. As the modulation bandwidth increases, the influence of spatial smoothing (modified either by wavelength detuning or by drift orthogonal to the IAWs) on CBET becomes progressively weaker. Because of the finite frequency range of the different laser beam wavelets, part of the ponderomotive grating components can be at the acoustic resonance even though the corresponding monochromatic (plane wave) case is out of resonance. This also indicates a broadening of the CBET resonance due to SSD. \\
Importantly, neglecting the dispersion of the SSD (thus accounting only for the frequency modulation) leads to an IAW grating without the stretching of Fig.~\ref{RA TSSD} and close to Fig.~\ref{RPP pondero acoustique vdy}. This indicates that operating SSD without the spatial dispersion does not mitigate the power exchange compared to the RPP case described in \cite{OudinPOP2}. 
Overall, this section demonstrates that incorporating temporal smoothing in CBET modeling is necessary regardless of the wavelength detuning between the beams or the direction of the plasma flow.

\section{\label{III Modele} Modeling CBET between spatially and temporally smoothed laser beams}
After qualitatively describing the influence of spatial and temporal smoothing of the beams on CBET in Part.~\ref{II Heuristique Interprétation}, we detail in subsections~\ref{III.1 Champs rpp ssd}, \ref{III.2 Battement pondéro}, \ref{III.3 Battement acoustique}, and \ref{III.4 Puissance échangée} the main assumptions and calculation steps leading to a simplified formulation of the power transfer between the smoothed beams. In subsection~\ref{III.5 Exchange disp et bandwidth}, we examine the influence of the beam modulation depth and the dispersive grating on the power transfer. In subsection~\ref{III.6 Influence synchronisation}, we analyze the impact of modulator desynchronization between the two laser chains.  \\

\subsection{\label{III.1 Champs rpp ssd} Field produced by a RPP and SSD beam } 
\begin{figure}
\centering
    \begin{subfigure}{0.4\textwidth}
        \caption{TSSD beam}
        \includegraphics[scale=0.4]{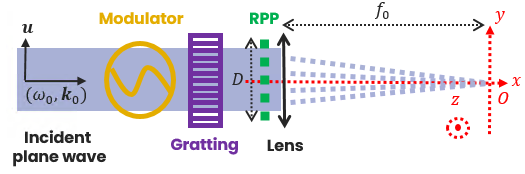}
        \label{TSSD}
    \end{subfigure}
    \begin{subfigure}{0.4\textwidth}
        \caption{LSSD beam}
        \includegraphics[scale=0.4]{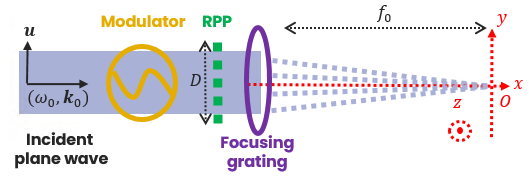}
        \label{LSSD}
    \end{subfigure}
    \vspace{-1.0em}
\caption{\colorbox{Gray!20}{\parbox{\dimexpr\linewidth-2\fboxsep}{\textbf{FIG \thefigure:} Simplified scheme of the temporal smoothing applied on the laser chain. Spatial smoothing is provided by a random phase plate (RPP, green). Temporal smoothing is achieved through the combination of a frequency modulator (orange) and spectral dispersion, implemented using a dispersive grating for transverse smoothing and a focusing grating for longitudinal smoothing (both shown in purple in panels (a) and (b), respectively). In the case of transverse smoothing, the temporally and spatially out-of-phase wavelets are focused by a converging lens (black). The reference frame $(Oxyz)$ is defined, with the transverse plane $(Oyz)$ centered on the RPP and the $(Ox)$ axis oriented orthogonally to it.}}}
\label{1 laser}
\end{figure}

We consider a $\mathcal{D}$-dimensional square RPP of side length $D$. For $\mathcal{D} = 1$, the RPP lies in the $(Oy)$ plane and consists of $N$ segments of length $d$, each imposing a random phase shift $\phi_{\mathbf{n}} \in [0, 2\pi]$, indexed by $\mathbf{n} = (0, n_y)$ with $n_y = 1, \dots, N$. For $\mathcal{D} = 2$, the RPP lies in the $(Oyz)$ plane and contains $N^2$ square elements of area $d^2$, each with a phase shift $\phi_{\mathbf{n}} \in [0, 2\pi]$, indexed by $\mathbf{n} = (0, n_y, n_z)$ where $n_y, n_z = 1, \dots, N$. 
The laser beam propagates along the $x$-axis, perpendicular to the RPP, with the origin of the axis set at the focal point, located at a distance $f_0$ from the RPP.
We associate the direct orthonormal basis \(\hat{\mathbf{e}}_x, \hat{\mathbf{e}}_y, \hat{\mathbf{e}}_z\) with the coordinate system \(Oxyz\). 
Before any smoothing is applied, the beam is modeled as a plane wave, linearly polarized along $\mathbf{u} = \hat{\mathbf{e}}_y$ for P polarization or $\mathbf{u} = \hat{\mathbf{e}}_z$ for S polarization, with wave vector $\mathbf{k}_0 = k_0 \hat{\mathbf{e}}_x$ in the plasma, angular frequency $\omega_0$, and group velocity $v_g = k_0 c^2 / \omega_0$ in the plasma.
We define \( \tilde{k} = k_0 / (2f_{\#}) \) as the maximum transverse wave vector imposed by the focusing optics, with \( f_{\#} = f_0 / D \) being the f-number, given by the ratio of the focal length \( f_0 \) to the total transverse extent \( D \) of the RPP. 
The RPP transforms the incident wave vector \( \mathbf{k}_0 \) into a discrete set of transverse wave vectors
\( \mathbf{k}_{\mathbf{n}} = \tilde{k} \frac{2\mathbf{n} - N - 1}{N},\)
which represents the angular spread of the beam due to diffraction. For \(\mathcal{D} = 1\), the wave vectors reduce to \( \mathbf{k}_{\mathbf{n}} = (0, k_{n_y}) \), and for \(\mathcal{D} = 2\),  \( \mathbf{k}_{\mathbf{n}} = (0, k_{n_y}, k_{n_z}) \). \\
Before passing through the RPP, the beams are modulated at frequency \( \nu_d = \omega_d / (2\pi) \), producing a frequency comb centered on \( \nu_0 \) with spectral width \( \Delta \nu_d \). We define $M = \Delta \nu_d / (2\nu_d)$ as the modulation depth of the beam. For a frequency tripled wavelength, $M$ corresponds \DB{to} three times the amplitude of the sinusoidal phase modulation applied to the beam by the modulator. This collection of multicolor wavelets is then dispersed by a grating in the case of transverse smoothing by spectral dispersion or by a focusing grating in the case of longitudinal smoothing by spectral dispersion, as illustrated in Fig.~\ref{TSSD} and Fig.~\ref{LSSD}, respectively. In the TSSD configuration, all wavelets pass through an achromatic lens with focal length \( f_0 \). In contrast, in the LSSD configuration, each wavelet experiences a different focal length~\cite{Bor:89}, and only the central frequency component is focused at a distance \( f_0 \).
TSSD and LSSD therefore couple wavelets that are phase-shifted in time by the modulator and in space by the RPP. For TSSD, a dispersive grating induces focusing of the wavelets at different positions along the transverse axis \( (Oy) \), while for LSSD, a focusing grating causes focusing at different positions along the longitudinal axis \( (Ox) \).
The envelope approximation is used to write the total field from the optical chain as $\textbf{A}(\textbf{r},t)=\tilde{\textbf{A}}(\textbf{r},t)e^{i[k_{0}x-\omega t]}$ where $\textbf{A}$ is the potential vector and $\tilde{\textbf{A}}$ is its slowly varying envelope~\cite{OudinPOP1}.
\DB{After passing through these different optical elements, the field near the focal plane is}~\cite{Huller,charles_2025,RuyerSBS,RuyerSpray} 
\begin{align}
    \frac{\tilde{\textbf{A}}(\textbf{r},t)}{A_{0}}=&\frac{\bf{u}}{N^{\frac{\mathcal{D}}{2}}}\sum_{\textbf{n}} e^{i\big\{\phi_{\textbf{n}}+\textbf{k}_{\textbf{n}}.\textbf{r}+M\sin(\psi^{\text{T/L}}_{\textbf{n}}-\omega_dt)\big\}} \ .
    \label{champ ref rpp}
\end{align}
The value of  $\vert \tilde{\textbf{A}}\vert ^2$ averaged over the phase plate random variables is independent of time and is $A_{0}^2$.
We neglect the spatial phase term proportional to $\omega_d x / c$ that would arise from the propagation of the modulated field. The frequency-dependent angular dispersion is captured by the phase term $\psi^{\mathrm{T/L}}_{\mathbf{n}}$.
For transverse SSD, this term is~\cite{RuyerSpray,RuyerSBS}
\begin{equation}
\psi^{\mathrm{T}}_{\mathbf{n}} = N_c \, \frac{2\pi f_\#}{k_0} \, \mathbf{k}_{\mathbf{n}} \cdot \hat{\mathbf{e}}_y,
\end{equation}
and for longitudinal SSD, it is~\cite{RuyerSpray,RuyerSBS}
\begin{equation}
\psi^{\mathrm{L}}_{\mathbf{n}} = N_c \, \frac{f_0}{2k_0} \, \frac{\omega_d}{\omega_0}\mathbf{k}_{\mathbf{n}}^2.
\end{equation}
The parameter $N_c = \nu_d T_r$ represents the total number of grating cycles accumulated across the transverse extent of the beam, where $\nu_d$ is the modulation frequency and $T_r$ the time delay introduced by the dispersive system. For LSSD, this delay is given by $T_r = D / (c f_\#)$, while for TSSD it is $T_r = \lambda_0 D / (c \, \delta \cos \theta_r)$, with $\delta$ the grating period and $\theta_r$ the diffraction angle of the central wavelength.
Under the paraxial approximation, this time delay is equivalent to a spatial shift of the modulated wavelets by approximately $N_c$ speckle size. For TSSD, this transverse shift is $l_s = N_cf_\# \lambda_0$, and for LSSD, the longitudinal shift is $L_s = 10 N_cf_\#^2 \lambda_0$. To match experimental conditions and ensure efficient smoothing, we set $N_c \approx 1$ in the simulations and numerical applications (except for Fig. \ref{Cycle number influence}). The phase modulation is expanded using cylindrical Bessel functions of the first kind as
\begin{equation}
e^{iM \sin\left[\Psi^{\mathrm{T/L}}_{\mathbf{n}} - \omega_d t\right]} \approx \sum_{m=-\Upsilon}^{+\Upsilon} J_m(M) \, e^{i m \left[\Psi^{\mathrm{T/L}}_{\mathbf{n}} - \omega_d t\right]} \ .
\end{equation}
We introduced the parameter \(\Upsilon\),  the truncation of the sum introduced into the spectrum of the modulated beam in order to simplify the numerical evaluation. 
$\Upsilon $ is typically chosen as a multiple of the modulation width ceiling, \(\lceil M \rceil\). 
For clarity, we set \(\Upsilon = \lceil M \rceil\) in the following, noting that this artificial truncation may lead to a slight underestimate of the laser beam power, on the order of a few percent.
Henceforth, all physical quantities related to the optical smoothing of the beams are considered at \(3\omega\).
Thus, the temporal spectrum of each laser beam considered in the following is reduced to $2\Upsilon+1$ frequencies, which corresponds to a bandwidth (in frequency) equal to $\Delta \nu_d=2\Upsilon\nu_d$. At the LMJ and the NIF, we use modulation depths of \( M = 15 \) and \( M = 3.9 \), respectively, and a modulation frequency of \( \nu_d = 14.25\,\text{GHz} \) and \( \nu_d = 17\,\text{GHz} \), respectively. This corresponds to a bandwidth of \( \Delta \nu_d = 427.5\,\text{GHz} \) for LMJ and \( \Delta \nu_d = 132.6\,\text{GHz} \) for NIF. 
In Eq.\eqref{champ ref rpp} it is clear that the total field is the superposition of wavelets phase-shifted in space by the phase plate, in time by the SSD and amplitude modulated by the Bessel functions which will distribute the laser beam energy on either side of the central frequency $\nu_0$ of the spectrum~\cite{Fusaro}, thus reducing efficiently the coupling between the laser beam and the plasma.

\subsection{\label{III.2 Battement pondéro} Ponderomotive grating induced by crossing two smoothed laser beams }
\begin{figure}
    \centering
    \includegraphics[scale=0.51]{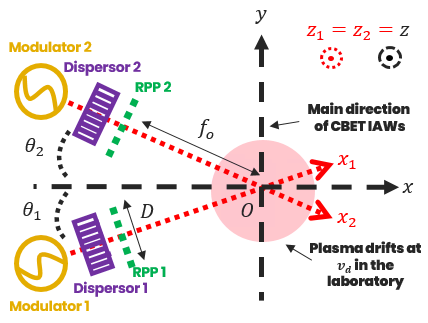}
    \caption{\colorbox{Gray!20}{\parbox{\dimexpr\linewidth-2\fboxsep}{\textbf{FIG \thefigure:} Schematic of the spatially and TSSD-smoothed beam crossing configuration in a plasma drifting in the $(xOy)$ plane of the laboratory. The two beams ($i=1,2$) propagate along the $(Ox_i)$ directions (red dashed arrows) and are inclined by an angle $\theta$ with respect to the longitudinal $(Ox)$ direction of the laboratory frame $(Oxyz)$ (black dashed arrows). Plasma moves at $\textbf{v}_d$ in the laboratory (light red).
    }}}
    \label{CBET scheme}
\end{figure}
We consider the crossing of two laser beams, each originating from a separate laser chain as shown in Fig.~\ref{1 laser}. In this subsection, the coordinate system \( (Oxyz) \) refers to the laboratory frame, as shown in Fig.~\ref{CBET scheme}, and not to the frame associated with a single laser chain, as discussed in Subsec.~\ref{III.1 Champs rpp ssd}. Both beams share the same modulation depth \( M \) and modulation frequency \( \nu_d \). The lower beam (beam~1) is inclined at an angle \( +\theta \), and the upper beam (beam~2) at \( -\theta \) with respect to the laboratory frame, as illustrated in Fig.~\ref{CBET scheme}. Before modulation, the beams have wave vectors and angular frequencies denoted by \( (\mathbf{k}_{0_1}, \omega_1) \) and \( (\mathbf{k}_{0_2}, \omega_2) \), respectively.  The two beams are linearly polarized along directions \( \textbf{u}_1 \) and \( \textbf{u}_2 \). In the case where the beams are linearly S polarized, then \( \mathbf{u}_1 \cdot \mathbf{u}_2 = 1 \). For linear P polarization the relative polarization angle is \( \theta \), as \( \mathbf{u}_1 \cdot \mathbf{u}_2 = \cos(2\theta) \). After passing through the modulators (indexed by \( m_1 \) and \( m_2 \)), dispersive elements (associated with the phase shifts \( \Psi^\mathrm{T/L}_{\mathbf{n}_1} \) and \( \Psi^\mathrm{T/L}_{\mathbf{n}_2} \)), and random phase plates (indexed by \( \mathbf{n}_1 \) and \( \mathbf{n}_2 \)), each beam splits into \( N^{\mathcal{D}}(2M + 1) \) wavelets that are randomly phase-shifted and intersect in the plasma on a pairwise basis.
The ponderomotive force includes a beating term of the form \( \mathbf{F}_p = e^2 / (2m_e) \nabla^2 (\mathbf{A}_1 \cdot \mathbf{A}_2^*) \). Within the envelope approximation, this product is expressed as \( \mathbf{A}_1 \cdot \mathbf{A}_2^* = \tilde{\mathbf{A}}_1 \cdot \tilde{\mathbf{A}}_2^* \, e^{i[\mathbf{k} \cdot \mathbf{r} - \omega t]} \), where the resulting wave vector and frequency are \( \mathbf{k} = \mathbf{k}_{0_1} - \mathbf{k}_{0_2} \simeq 2k_0 \sin(\theta) \, \hat{\mathbf{e}}_y \) and \( \omega = \omega_1 - \omega_2 \). These correspond to the interference pattern formed by the two crossing plane waves~\cite{Debayle_op}.
The wave vector associated with each ponderomotive beat, resulting from the coupling of two spatially and temporally smoothed beams, is defined as
\begin{align}
    &\mathbf{K}_{\mathbf{n}_1\mathbf{n}_2}
    = \mathbf{K}_{\mathbf{n}_1} - \mathbf{K}_{\mathbf{n}_2} \nonumber 
    = \bigl\{\sin(\theta)\,[k_{n_{y_1}} + k_{n_{y_2}}]\bigr\}\,\hat{\mathbf{e}}_x \\
    &+ \bigl\{k - \cos(\theta)\,[k_{n_{y_1}} - k_{n_{y_2}}]\bigr\}\,\hat{\mathbf{e}}_y
    + \bigl\{k_{n_{z_1}} - k_{n_{z_2}}\bigr\}\,\hat{\mathbf{e}}_z \ .
    \label{Wave vector}
\end{align}
Since the influence of SSD on the wave vector arising from the beating of the two beams in Eq.~\eqref{Wave vector} is negligible, we thus consider this wave vector to depend solely on the phase plates. However, the phase modulation of the SSD strongly modifies the angular frequency of these wavelets. The total angular frequency of each beat is expressed as
\begin{equation}
    \Omega_{m_{1}m_{2}} = \omega_1 - \omega_2 + [m_1 - m_2]\omega_d \ .
\end{equation}
The ponderomotive grating imposed on the plasma is given by~\cite{charles_2025}
\begin{align}
    \frac{\textbf{A}_1 \cdot \textbf{A}_2^*}{A_{1,0} A_{2,0}^*} 
    &= \frac{\textbf{u}_1 \cdot \textbf{u}_2}{N^{\mathcal{D}}} 
    \sum_{\textbf{n}_{1}, m_{1}} \sum_{\textbf{n}_{2}, m_{2}} 
    \prod_{j=1}^{2} J_{m_j}(M) \, 
    e^{i\left\{\phi_{\textbf{n}_1} - \phi_{\textbf{n}_2}\right\}} \nonumber\\
    &\quad \times e^{i\left\{ \textbf{K}_{\textbf{n}_{1}\textbf{n}_{2}} \cdot \textbf{r} 
    - \Omega_{m_{1} m_{2}} t 
    + m_1 \left[ \psi^{\text{T/L}}_{\textbf{n}_1} - \omega_d t_0 \right] 
    - m_2 \psi^{\text{T/L}}_{\textbf{n}_2} \right\}} \ .
    \label{pondero beat}
\end{align}
In this expression, the beam 1 modulator is assumed to be time-delayed by \( t_0 \) with respect to the upper modulator. This is equivalent to replacing \( t \) with \( t + t_0 \) in Eq.~\eqref{champ ref rpp}, which describes the field induced by the lower beam. The influence of this time delay, will be studied in Subsec. \ref{III.6 Influence synchronisation}. 

\subsection{\label{III.3 Battement acoustique} Plasma acoustic response: Linear kinetic model}
We consider a homogeneous plasma at initial density $n_{e_0}$, at plasma frequency $\omega_{pe}=\sqrt{e^2n_{e_0}/(\epsilon_0m_e)}$, where $m_e$ and $e$ are the mass and charge of the electrons. The plasma is composed of electrons and several species of ions indexed by $p$. The number of nucleons and the atomic number of the nucleus $p$ are noted $A_p$ and $Z_p$.
The plasma response is derived using linear kinetic theory~\cite{Drake,Michel_POP_2009,OudinPOP2} within the envelope approximation. The plasma species drift with velocity 
\(\mathbf{v}_d = (v_{dx}, v_{dy}, v_{dz})\) in the laboratory frame (as illustrated in Fig.~\ref{CBET scheme}), 
and their density is linearized as \(n_p = n_{p_0} + \delta n_p\). 
The density fluctuations driven by the ponderomotive grating~\cite{charles_2025,RuyerSBS,RuyerSpray,OudinPOP2} are
\begin{align}
    \frac{\delta n_e}{n_{e_0}}&=\frac{\textbf{u}_1\cdot\textbf{u}_2}{N^{\mathcal{D}}}\frac{A_{0,1}A_{0,2}^*\epsilon_0}{2m_en_{e_{0}}}\sum_{\textbf{n}_{1},m_{1}}\sum_{\textbf{n}_{2},m_{2}}\prod_{j=1}^{2} J_{m_j}(M)\nonumber\\&\times e^{i\big\{\phi_{\textbf{n}_1}-\phi_{\textbf{n}_2}-\Omega_{m_{1}m_{2}}t+\textbf{K}_{\textbf{n}_{1}\textbf{n}_{2}}.\textbf{r}\big\}} f(\Omega_{m_1m_2},\textbf{K}_{\textbf{n}_{1}\textbf{n}_{2}}) \nonumber \\& \times e^{ i\big\{m_1[\psi^{\text{T/L}}_{\textbf{n}_1}-\omega_dt_0]-m_2\psi^{\text{T/L}}_{\textbf{n}_2}\big\}} \ .
    \label{dne}
\end{align} 
The linear kinetic response of the plasma is given by~\cite{Drake,Michel,Debayle_rt}
\begin{equation}
    f(\Omega, \textbf{K}) = \textbf{K}^2 \, 
    \frac{\chi_e(\Omega, \textbf{K}) \left[ 1 + \sum_p \chi_p(\Omega, \textbf{K}) \right]}
         {1 + \chi_e(\Omega, \textbf{K}) + \sum_p \chi_p(\Omega, \textbf{K})} \ .
    \label{CBET peak kin}
\end{equation}
Here, $\chi_p$ represents the dielectric susceptibility tensor of species $p$, 
\begin{equation}
    \chi_{p}(\Omega,\textbf{K})=-\frac{1}{2\textbf{K}^2\lambda_{p}^2}Z'\left(\frac{\Omega -\textbf{K} \cdot \textbf{v}_d }{\sqrt{2}|\textbf{K}|v_{th_{p}}}\right) \, ,
\end{equation}
where $v_{th_p}=\sqrt{T_p/m_p}$ denotes the thermal velocity of species $p$ and $Z'$ is the derivative of the plasma dispersion function \cite{Fried_1960}. \\
The linear fluid limit of the CBET resonance peak~\cite{OudinPRL,OudinPOP2}, given in Eq.~\eqref{CBET peak kin}, can be expressed as
\begin{equation}
    f(\Omega, \mathbf{K}) = 
    \frac{\left( \frac{m_e \, \omega_{pe}^2}{m_i \, c_s^2} \right) K^2 \, c_s^2}
         {K^2 \, c_s^2 - (\Omega - \mathbf{K} \cdot \mathbf{v}_d)^2 - 2 i \, \nu_{\text{L}} (\Omega - \mathbf{K} \cdot \mathbf{v}_d)} \,,
    \label{eq:fluid}
\end{equation}
where $\nu_{\text{L}}$ is the linear Landau damping rate, whose value is obtained following the procedure outlined in~\cite{OudinPOP2}.

\subsection{\label{III.4 Puissance échangée} Main assumptions for a simplified steady-state power exchange formulation}
Within the linear theory, the vector potential, and intensity of beam \(i\) are expressed as
\begin{equation}
    \mathbf{A}_{i} = \mathbf{A}_{i}^0 + \delta \mathbf{A}_{i}, \quad
    I_{i} = I_{i}^0 + \delta I_{i}.
\end{equation}
The initial intensity (assumed identical for both beams) and its linear perturbation are given by
\begin{equation}
    I_{i}^0 = A_{i,0}^2 \epsilon_0 \omega_0^2 v_g / 2, \quad
    \delta I_i = \mathrm{Re} \{ A_i^0 \delta A_i^{*} \} \, \epsilon_0 \omega_0^2 v_g.
\end{equation}
In the following, all quantities are expressed in terms of the linear power density \(P_i = I_i L\), where \(L\) represents the transverse width of the beam-crossing region, defined as
\begin{equation}
    L = N f_{\#} \lambda_0 = R \sin(2\theta) \,.
\end{equation}
To simplify, we introduce the power difference between the beams and the total power as
\begin{equation}
    \delta P = \delta P_1 - \delta P_2, \quad P = P_1^0 + P_2^0 \,.
\end{equation}
Assuming \(\delta P_1 = - \delta P_2\) and \(P_1^0 = P_2^0\), the relative power variation can be written as
\begin{equation}
    \frac{\delta P}{P} = \frac{\delta P_1}{P_1^0} = - \frac{\delta P_2}{P_2^0} \,.
\end{equation}
The exchanged intensity between the beams is calculated within the framework of asymptotic linear perturbation theory by integrating the laser beam field transport equation, Eq.~\eqref{Eq transport}, over the longitudinal extent of the beam-crossing region, \(R\). We then integrate the relative intensity variations between the beams over the transverse crossing region of extent \(L^D\), yielding an expression for the relative power variation \(\delta P / P\) (Eq.~\eqref{moy y0}). The details of these calculations are provided in Appendix~\ref{app3 Derivation de l'échange}.
Note that the spatial shift, on the order of \( Mf_{\#} \lambda_0 \), between the different colors is neglected here. This approximation is true for large beams, but probably more questionable for small beams such as those used in PIC simulations.
This expression depends both on the phase plate configuration and time. Since the RPP elements are fully decorrelated~\cite{OudinPOP2}, we perform an average over an infinite number of phase plate realizations, obtaining, for SSD beams, the temporal fluctuation of power exchanged between the beams \(\langle \delta P(t)/P \rangle_\phi\) (Eq.~\eqref{fluctuation t}). For spatially and temporally smoothed beams, this quantity varies with time, in contrast to cases considering only spatial smoothing.
We then choose to subsequently reduce the power exchange to its time average, thus neglecting the impact of the temporal fluctuations.
We then  average over one modulation period to obtain the time-averaged power exchange as detailed in Appendix~\ref{app3 Derivation de l'échange}.
We obtain, 
\begin{align}
    \frac{\langle \delta P \rangle_\phi^t}{P} = &  \frac{-P_2^0 C}{\sin(2\theta) N^{2\mathcal{D}}} \sum_{\mathbf{n}_1,m_1} \sum_{\mathbf{n}_2,m_2} \sum_{m_3} \prod_{j=1}^3 J_{m_j}(M) \nonumber \\
    & \times J_{m_2 + m_3 - m_1}(M) \, f(\Omega_{m_1 m_2}, \mathbf{K}_{\mathbf{n}_1 \mathbf{n}_2}) \nonumber \\
    & \times e^{i \left( \psi^{\mathrm{T/L}}_{\mathbf{n}_1} - \psi^{\mathrm{T/L}}_{\mathbf{n}_2} - \omega_d t_0 \right)(m_1 - m_3)},
    \label{mean t}
\end{align}
where $C = \frac{(\mathbf{u}_1 \cdot \mathbf{u}_2)^2}{2 \epsilon_0 \omega_0^3 v_g^2} \frac{e^2}{m_e^2}$. Here, $v_g$ is the laser group velocity.\\
A simpler power exchange formula results from  averaging Eq.~\eqref{mean t} over $t_0$ which could be relevant when $t_0$ is not well constrained or for efficient numerical evaluation. This leads to a term $\mathrm{sinc}(\pi (m_1 - m_3)) \approx \delta_{m_1,\,m_3}$, where $\delta_{i,\,j}$ denotes the Kronecker delta, which allows the expression to be simplified as
\begin{align}
    \frac{\langle \delta P \rangle_{\phi, t_0}^{t}}{P} 
    &= \frac{-P_2^0 \, C}{\sin(2\theta) \, N^{2\mathcal{D}}} 
       \sum_{\textbf{n}_1, m_1} \sum_{\textbf{n}_2, m_2} 
       \prod_{j=1}^2 J_{m_j}^2(M) \nonumber\\
    &\quad \times f(\Omega_{m_1 m_2}, \textbf{K}_{\textbf{n}_1 \textbf{n}_2}) \,.
    \label{mean desynchro}
\end{align}
It is important to note that the grating dispersion term, $\psi_{\textbf{n}}^{\text{T/L}}$, disappears in Eq.~\eqref{mean desynchro} compared to Eq.~\eqref{mean t}.
This implies that when $v_{dz}=0$, the sum over $n_{1z}$ and $n_{2z}$ becomes trivial, $N^{-2} \sum_{n_{1z},n_{2z}} = 1$,  and the three-dimensional power exchange coincides with its 2D counterpart. Finally, accounting for PS simply consists in replacing $(\mathbf{u}_1\cdot \mathbf{u}_2)^2$ by $1/2$ \cite{Michel_POP_2009}. 

\subsection{\label{III.5 Exchange disp et bandwidth}Influence of SSD dispersion and beam bandwidth on CBET}

\begin{figure}
\centering
\begin{subfigure}{0.48\textwidth}
    \centering
    \caption{C plasma, $N_c=1$}\vspace{-0.4em}
    \includegraphics[scale=0.32]{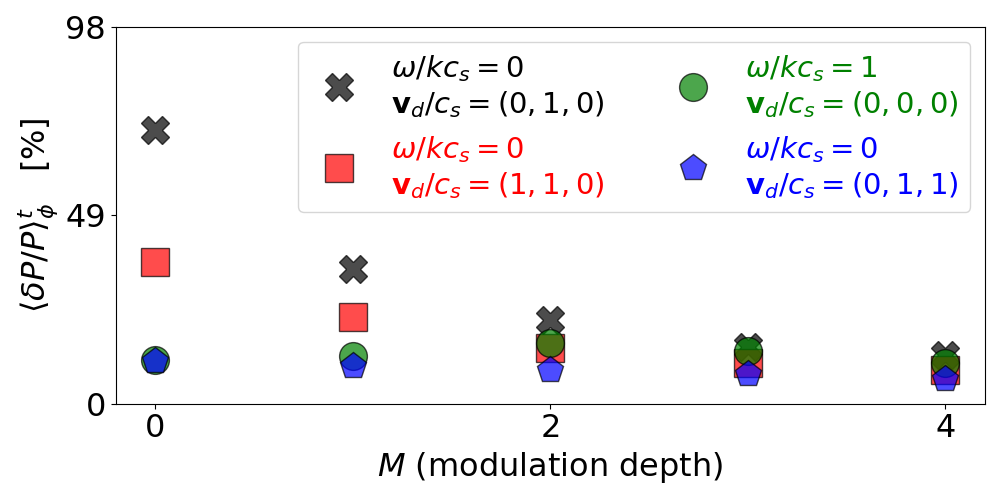}
    \label{bandwidth influence on CBET}
\end{subfigure} 
\begin{subfigure}{0.48\textwidth}
    \centering
    \caption{CH plasma}\vspace{-0.4em}
    \includegraphics[scale=0.32]{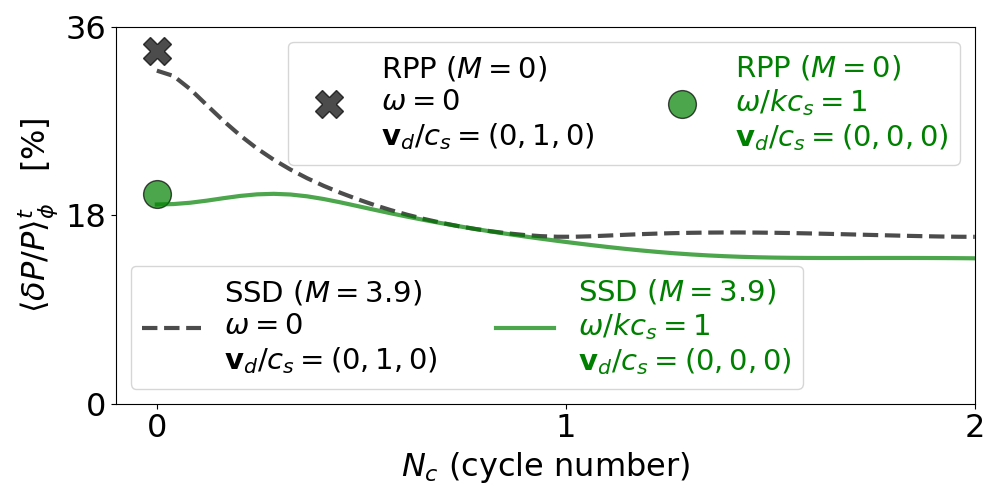}
    \label{Cycle number influence}
\end{subfigure}
\vspace{-1.0em}
\caption{\colorbox{Gray!20}{\parbox{\dimexpr\linewidth-2\fboxsep}{\textbf{FIG \thefigure:} Power exchange between two NIF-type TSSD beams as predicted by Eq.~\eqref{mean t} as a function of the modulation depth $M$ for $N_c=1$ (a) for a C plasma, and as a function of cycle number $N_c$ for $M=3.9$ in (b) for a CH plasma. (a) Crosses, squares, circles, and pentagons indicate the power transfer induced by a $y$-aligned flow, frequency shift, $xOy$-aligned flow, and $yOz$-aligned flow, respectively. 
(b) Power exchange from Eq.~\eqref{mean t} in the frequency shift (solid green line) and the drift case (dashed black line) for $M=3.9$. 
Markers show the results obtained without SSD. Except for $N = 21$, $\mathcal{D}=2$, $P_{i,\text{CH}}^0 = 70\,\text{TW·m}^{-1}$ and $P_{i,\text{C}}^0 = 30\,\text{TW·m}^{-1}$, all other parameters are the same as in Fig.~\ref{TSSD pondero acoustique vdy}.}}}
\label{Dispersion and Bandwidth influence}

\end{figure}

Figure~\ref{bandwidth influence on CBET} illustrates the power exchange in the fluid limit between two spatially and temporally smoothed beams, averaged over time and RPP configurations, $\langle \delta P / P \rangle_\phi^t$ from Eq.~\eqref{mean t} (markers), as a function of the modulation depth \( M \) for a two-dimensional RPP (\(\mathcal{D} = 2\)).
Black crosses, red squares, green circles, and blue pentagons correspond to $y$-aligned flow, wavelength shift, $xOy$-aligned flow, and $zOy$-aligned flow, respectively. When the flow is perturbed along a direction other than $(Oy)$, we consider a near-sonic flow along both directions. The NIF parameters are used in Figs.~\ref{bandwidth influence on CBET} for a carbon plasma and in Figs.~\ref{Cycle number influence} for a CH plasma with the cycle number fixed to $N_c=1$. The case $M=0$ corresponds to purely spatially smoothed beams.
Figure~\ref{bandwidth influence on CBET} clearly demonstrates that increasing the SSD modulation depth significantly reduces the average power exchanged at resonance compared to the case with only spatial smoothing ($M=0$), regardless of plasma drift or wavelength detuning. 
We nevertheless observe a slight increase in the exchanged power at low modulation depth for detuned beams. This behavior was already identified in earlier work on SBS~\cite{charles_2025}. 
%
We also shows that the effects of a wavelength shift, a $v_{dx}$ drift, and a $v_{dz}$ drift are pronounced for two crossing RPP beams but become negligible as the modulation depth increases. 
%
For SSD beams, the influence of $\omega$ and $v_{dx}$ is generally subdominant, but it cannot be neglected in very weakly damped plasmas, such as gold or if the modulation depth is not large enough. Note that the position of the resonance  $\omega -kv_{dy}=kc_s$ remains unaffected by the laser beam smoothing. Compared to the usual plane wave predictions and under these conditions, deviations of approximately 10\% for the wavelength shift, 20\% for drift in the $(xOy)$ plane, and 50\% in the $(yOz)$ plane can be expected, particularly at the NIF, where the number of spectral components remains limited.
The significant impact of SSD on CBET has not been fully appreciated in previous works, because the spatial dispersion of the SSD is often neglected. \\
Figure~\ref{Cycle number influence} shows the average power exchanged between the beams as a function of the number of cycles $N_c$ and for $M=3.9$, highlighting the importance of grating-induced spatial dispersion. This power exchange is evaluated at acoustic resonance for beams that are both spatially and temporally smoothed. Two situations are considered: a drifting plasma (dashed black curve for SSD beams and cross markers for RPP beams), and beams with a wavelength shift (solid green curve for SSD beams and circular markers for RPP beams).
When SSD is applied without spatial dispersion, spatial smoothing dominates the power exchange, as evidenced by the close agreement at $N_c=0$ between the SSD curve and the markers (no bandwidth) in Fig.~\ref{Cycle number influence}. The small difference is related to the minimalist truncation of the laser beam spectrum, \(\Upsilon = \lceil M \rceil\) , as detailed in Subsec.~\ref{III.1 Champs rpp ssd}. In this regime, the beam aperture fully governs the CBET process, and the wavelength shift has a dominant influence, as demonstrated in~\cite{OudinPOP2}. By neglecting grating dispersion~\cite{BATES2020100772,Michel,Seaton1,Seaton2}, one assumes that all spectral components are focused at the same position $(x=0, y=0, z=0)$, so that the IAW and grating envelopes coincide, maximizing the power exchange. For the modulation bandwidths currently used in high-energy facilities (NIF, LMJ, Omega), SSD without spatial dispersion has no significant effect on power exchange.
When spatial dispersion is considered in addition to frequency modulation, as in experimental conditions, the maxima of the acoustic grating (Fig.~\ref{RA TSSD}) are stretched relative to the ponderomotive grating maxima (Fig.~\ref{RP TSSD}), which reduces CBET near acoustic resonance, as shown in Fig.~\ref{bandwidth influence on CBET}.
Figure~\ref{Cycle number influence} demonstrates that the number of cycles $N_c$ is an important parameter for CBET. However, shifting the focal points of the modulated spectral components by more than one speckle size (i.e., using $N_c>1$) provides no further stabilization of the power exchanged. In ICF experiments, the number of cycles is typically set to unity ($N_c \approx 1$).

\subsection{\label{III.6 Influence synchronisation}Influence of modulator desynchronisation on CBET}
\begin{figure}
\centering
\includegraphics[scale=0.35]{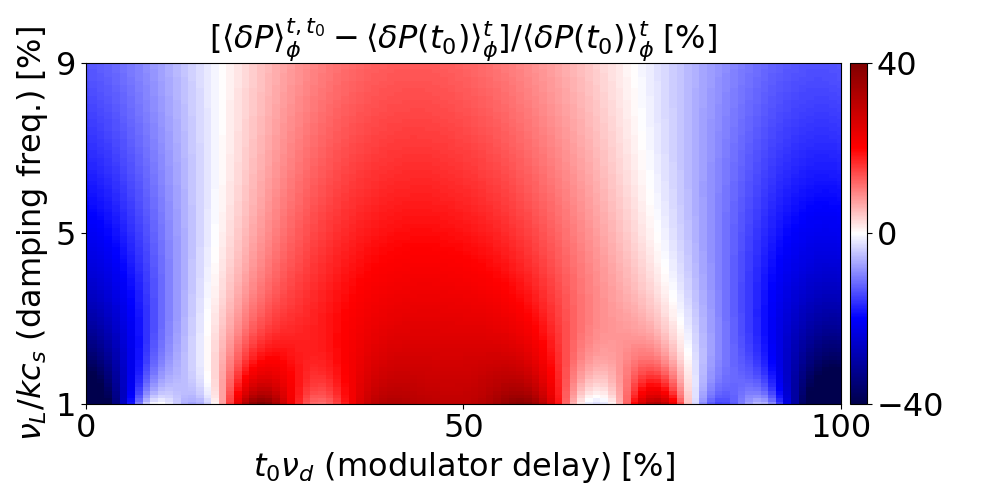}
\vspace{-0.6em}
\caption{\colorbox{Gray!20}{\parbox{\dimexpr\linewidth-2\fboxsep}{\textbf{FIG \thefigure:} Map of the normalized relative difference in power exchanged between Eq.~\eqref{mean desynchro} and Eq.~\eqref{mean t} for TSSD beams, as a function of the delay between the modulators of the two laser chains normalized to the modulation period $t_0\nu_d$ and the linear Landau damping rate normalized to the ion-acoustic frequency $\nu_L/kc_s$. We recall that $\langle\cdot\rangle_\phi$, $\langle\cdot\rangle^{t}$, and $\langle\cdot\rangle^{t_0}$ denote, respectively, averaging of the power exchange over phase-plate realizations, with respect to the modulation time, and with respect to the relative SSD synchronization between the two laser chains. The parameters are identical to those used in Fig.~\ref{bandwidth influence on CBET}, except for $\theta = 7^\circ$, $N = 41$, $\mathcal{D} = 1$, and  $\lambda_1 - \lambda_2 = 1.8\,\text{\AA}$.}}}
\label{desynchro influence on CBET}
\end{figure} 
In this section, we investigate the impact of SSD desynchronization on CBET. As described in Sec.~\ref{III Modele}, we impose a time delay $t_0$ between the frequency modulators of the two laser chains, as expressed in Eq.~\eqref{mean t}. However, since this delay is often not precisely known (or not communicated) in large-scale facilities, we perform an average over $t_0$, which leads to Eq.~\eqref{mean desynchro}.  \\
Figure~\ref{desynchro influence on CBET} shows a map of the normalized relative deviation between the power exchange in the fluid limit for synchronized beams (as predicted by Eq.~\eqref{mean t}) and for desynchronized beams (averaged over the modulator delay $t_0$, as predicted by Eq.~\eqref{mean desynchro}), plotted as a function of the normalized delay $t_0\nu_d$ and the normalized Landau damping rate $\nu_L/kc_s$. This representation allows both an assessment of the error introduced by averaging over $t_0$ and an exploration of the desynchronization influence under different plasma conditions. The crossing angle is set to $2\theta = 14^\circ$. Figure~\ref{desynchro influence on CBET} corresponds to NIF-type beams ($M=3.9$, $\nu_d = 17\,\text{GHz}$, $\lambda_1 - \lambda_2 = 1.8\, \text{\AA}$). Note that the wavelength shift modifies the CBET resonance to $v_{d||IAW}/c_s = 0.41$. 
Figure~\ref{desynchro influence on CBET}  shows that the exchanged power is highly sensitive to this delay, particularly in weakly damped plasmas where the  exchange varies by $\pm 40\%$ compared to the $t_0$-averaged value. The exchanged power reaches its maximum at $t_0 = 0$ and $t_0 = t_d$, i.e., when the beams are synchronized. The modulator desynchronization ($t_0 \neq 0,\ t_d$) reduces the power exchanged between beams at the acoustic resonance.  
Averaging the power exchange over the SSD modulator delay eliminates the dispersion term in Eq.~\eqref{mean t}. However, the power exchange predicted by Eq.~\eqref{mean desynchro} is not equivalent to that predicted by Eq.~\eqref{mean t} with $N_c = 0$ (synchronized, dispersion-free SSD). Averaging over $t_0$ therefore produces an effective reduction of CBET (with or without dispersion), unlike the case of non-dispersed SSD, as discussed in Sec.~\ref{III.5 Exchange disp et bandwidth}. This procedure also recovers the formula given in~\cite{Michel}, thereby clarifying its range of validity. Specifically, we show that it does not apply to synchronized beams or to weakly desynchronized beams ($t_0 < 0.25 t_d$, see Fig.~\ref{desynchro influence on CBET}). Figure~\ref{desynchro influence on CBET} further demonstrates that the lower the Landau damping frequency, the less accurate this averaging becomes.  \\
However, in the general case where the delay between the two laser chains is known, both synchronization and dispersion must be explicitly accounted for to quantitatively predict the power exchanged between the beams by substituting the exact values of \(t_0\) and \(N_c\) into Eq.~\eqref{mean t}.\\
To understand the effect of laser chain synchronization on CBET, we consider the beating term \( \mathbf{A}_1 \mathbf{A}_2^* = \mathbf{A}_1^{\text{RPP}} \mathbf{A}_2^{*\text{RPP}} e^{-iM\left[\sin(\omega_m t) - \sin\left(\omega_m(t - t_0)\right)\right]} \), which describes the interaction of two spatially and temporally smoothed laser beams, while neglecting the spatial phase dispersion for simplicity. 
The field \(A_i^{\text{RPP}}\) corresponds to the spatially smoothed field of beam \(i\), given by Eq.~\eqref{champ ref rpp} with \(M = 0\).
When the SSD of the beams are synchronized, this beating is equivalent to that of two RPP beams due to the absence of phase dispersion imposed on the beams:  \( \mathbf{A}_1 \mathbf{A}_2^* = \mathbf{A}_1^{\text{RPP}} \mathbf{A}_2^{*\text{RPP}}\). However, when SSD of the beams are desynchronized, the beating can be recast as \( \mathbf{A}_1 \mathbf{A}_2^* = \mathbf{A}_1^{\text{RPP}} \mathbf{A}_2^{*\text{RPP}} e^{-i M_{\text{Eff}} \cos(\omega_m t - \omega_d t_0 / 2)} \), and is equivalent to the interaction between an RPP beam and an SSD beam, where the effective modulation depth is given by \( M_{\text{Eff}} = 2M \sin\left(\omega_d t_0/2 \right) \). 
Therefore, desynchronizing the modulators of the laser chains results in a ponderomotive grating with a frequency width of $ 4M \sin\left(\omega_d t_0/2 \right)\omega_d$ thus affecting the driven acoustic waves and the resulting power exchange depending on the value of $t_0$.

\section{\label{V Comparaison Hydro}Comparison against Paraxial-Hydrodynamic Simulation Results}
\begin{figure}
\centering
        \includegraphics[scale=0.32]{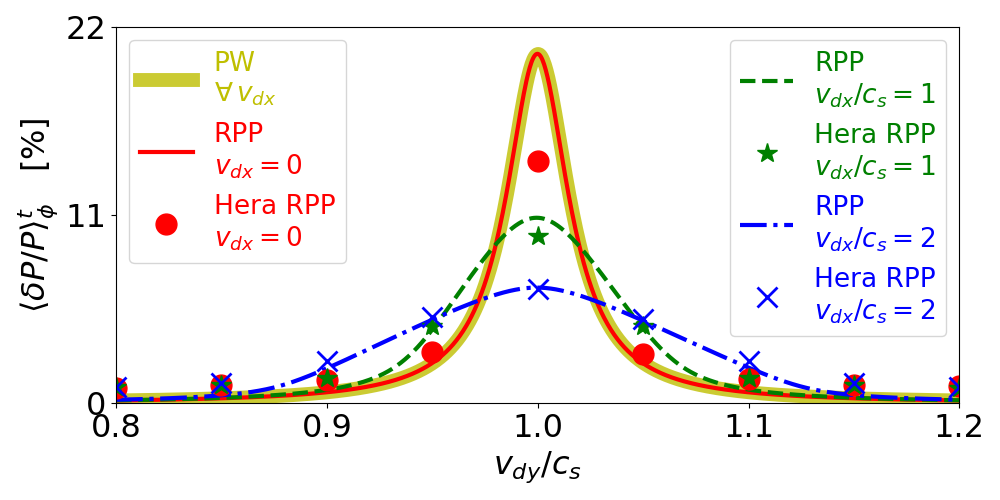}
        \vspace{-1.0em}
\caption{\colorbox{Gray!20}{\parbox{\dimexpr\linewidth-2\fboxsep}{\textbf{FIG \thefigure:}
Power exchanged between two synchronized TSSD beams in a carbon plasma, as predicted by Eq.~\eqref{mean t} (lines), extracted from the steady-state regime of our HERA  simulations as a function of  \( v_{dy} \). The average intensity \( I_i^0 = 2.17\,\rm  TW/cm^2 \), the sound speed  is \( c_s/c =1.15\times 10^{-3} \) and the IAW damping rate is $\nu/kc_s=0.018$.}}}
\label{cbet_hera}
\end{figure}

In this section, we compare the model presented in Sec.~\ref{III Modele} with a fluid plasma response against paraxial fluid simulations using Hera~\cite{Hera1,Hera2} (Figs.~\ref{cbet_hera}). The simulations are represented by markers, while the model by solid or plain lines. \\
Figure~\ref{cbet_hera} addresses the CBET resonance in the fluid simulations without SSD and for different orientations of the drift velocity. 
In the HERA simulations, we consider a carbon plasma composed of two species: electrons with density $n_{e_0} = 0.04\, n_c$ at temperature $T_e = 2\,\text{keV}$, and fully ionized carbon ions ($Z_{\mathrm{C}} = 6$, $A_{\mathrm{C}} = 12$) at temperature $T_i = 1\,\text{keV}$.
The average intensity of the S-polarized beams is $I_0 = 2.17\,\mathrm{TW/cm^2}$, the  wavelength is $\lambda_0=1\,\rm \mu m$ and the plasma has an acoustic damping rate of $\nu/kc_s=0.018$ that is  applied transversely to the laser beam direction through an operator evaluated in the the Fourier space~\cite{POP_Rose_96, Masson_2006}. In these simulations, the sound speed is $c_s/c = 1.15 \times 10^{-3}$, the cell size is $0.16 \times 0.007\, c^2 \omega_0^{-2}$, and the two beams intersect with a total crossing angle of $2\theta = 24^\circ$.
The value of $c_s$ is subsequently used to prescribe the plasma drift velocity in the HERA code.
The red line and markers correspond to the RPP case (no SSD) where the plasma drifts only in the $y$ direction. The theoretical predictions agree very well with the simulation results and coincide with the plane wave predictions (as a yellow line), as addressed in Ref. \cite{OudinPOP2}. The finite value of the $x$ component of the drift (in green and blue) broadens the resonance, and also in very good agreement with the predictions from HERA. The linearized framework in which the model is developed seems well verified in the HERA simulations until power exchange around approximately $15$--$20\%$. This broadening is more significant as  $v_{dx}$ becomes larger. \\

\section{\label{VI Comparaison PIC}Comparisons against PIC simulation results}
In this section, we compare the model predictions previously established in Sec.~\ref{III Modele} against Particle-In-Cell (PIC) simulations performed using the Smilei code. We first describe the parameters common to all simulations in Subsec.~\ref{V.1 Common para}, as well as the methodology used to compare the model with the simulations. In Subsec.~\ref{V.2 vdx}, within the context of spatially smoothed beam crossing, we validate the model’s ability to capture the effect of a plasma flow along \((Ox)\), transverse to the main IAW propagation direction \((Oy)\).
Next, we assess the impact of temporal smoothing on CBET in two distinct scenarios: one with a plasma drift along the IAW propagation direction but no wavelength shift between the beams (Subsec.~\ref{V.3 vdy}), and another with a wavelength shift but no plasma drift (Subsec.~\ref{V.4 w}).
This comparison between our linear model and the Smilei PIC code aims to validate the physical trends predicted by the linear theory. As detailed in Ref.~\cite{OudinPOP2}, our CBET model neglects the nonlinear effects (such as particle trapping, harmonic generation, and pump depletion) or the coupling with other linear effects, such as Brillouin backscattering. A precise understanding of the influence of these non-linear effects as done in Refs.~\cite{Nguyen,Seaton1,Yin_2023b}, and how they are affected by the laser beam smoothing requires a separate study and is therefore left for future work.
Importantly, our model neglects the effects related to the beam envelope. In particular, the dispersion used in the SSD shifts the focusing point of the different laser beam colours, as mentioned in Sec.~\ref{III.1 Champs rpp ssd}. This spatial shift should be negligible for large enough beams such as in ICF experiments and consistently with our analytical model. Because of the high numerical cost of the following simulations, the size of the beam injected in the plasma  is modest, around $\sim 48\,\rm \mu m$, and approaches the spatial shift induced by the maximum and minium frequency of the injected laser beam if one uses a large modulation depth. This restricts the PIC simulations to relatively modest modulation depths.


\subsection{\label{V.1 Common para} Common Simulation Parameters and Comparison Methodology}

\begin{figure}
\centering
    \begin{subfigure}{0.4\textwidth}
        \caption{SSD beams intensity profile}\vspace{-0.5em}
        \includegraphics[scale=0.40]{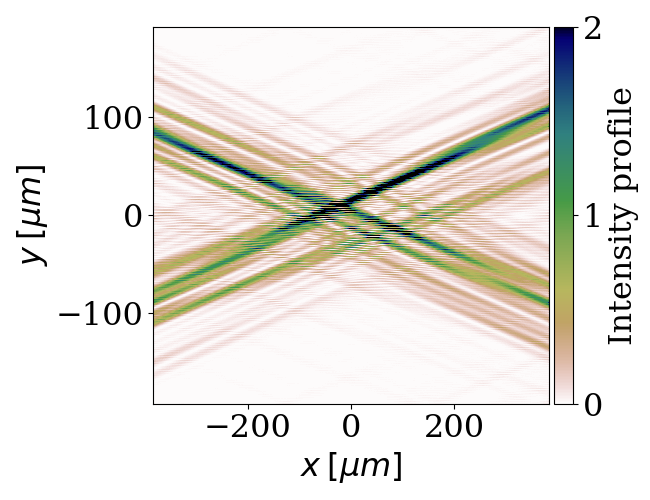}
        \label{I(x,y)}
    \end{subfigure}
    \hspace{0.25cm}
    \begin{subfigure}{0.4\textwidth}
        \caption{CBET time fluctuations}\vspace{-0.5em}
        \includegraphics[scale=0.43]{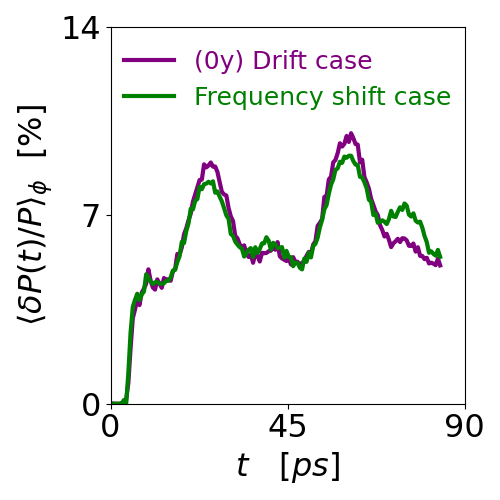}
        \label{I(t)}
    \end{subfigure}
\vspace{-1.0em}
\caption{\colorbox{Gray!20}{\parbox{\dimexpr\linewidth-2\fboxsep}{\textbf{FIG \thefigure:} Smilei PIC code predictions: (a) Intensity profile of the LSSD field normalized to $10^{17}$ W/m$^2$. (b) Temporal fluctuations of the power exchanged in the frequency shift case ($\omega=kc_s$, $\mathbf{v_d}=0$) and in the drift case ($\omega=kc_s$, $\mathbf{v_d}=c_s\Hat{\mathbf{e}}_y$). The plasma is made of CH with $n_e=0.04n_c$, $T_e=2\,\rm keV$ and LSSD synchronized beams are used ($t_0=0$, $\nu_d = 14.25$ GHz and $M = 2$).}}}
\label{Smilei presentation}
\end{figure}

This subsection outlines the parameters common to all PIC simulations and provides a brief description of the methodology used to compare the linear model against these simulations. 
The linear kinetic model developed in Sec.~\ref{III Modele} is validated here using the Smilei PIC code~\cite{Smilei}, where the laser beam is spatially and temporally smoothed by RPP and SSD, whose exact formulations are presented in Appendix~\ref{app2 Smilei model}. 
Note that the speckles have a finite longitudinal length in our simulations.
The two dimensional simulation box measures $4824\times2412 \, c^2\omega_0^{-2}$ and contains cells of dimension $0.4\times0.4 \, c^2\omega_0^{-2}$. Each cell consists of $300$ macro-electrons and macro-ions. The simulation time step is $0.04 \, \nu_0^{-1}$.
We consider a CH plasma with three  species, namely electrons with density $(n_{e_0}=0.04\:n_{c})$ at temperature $T_e=2 \,\text{keV}$, fully ionized carbon ($Z_{\text{C}}=6 \quad A_{\text{C}}=12$), and protons ($Z_{\text{H}}=1 \quad A_{\text{H}}=1$) in stoichiometric proportion. The initial ion temperature is set to $T_i = 1\,\text{keV}$.
The simulations are carried out without collisions.
The ion acoustic speed is determined by numerically solving the linear acoustic dispersion relation for the plasma’s free modes, as described in Ref.~\cite{OudinPOP2}. This yields an acoustic velocity of \( c_s/c = 1.09 \times 10^{-3} \). These values are subsequently employed to set the plasma drift velocity within the Smilei code. Additionally, the lasers wavelengths are hereafter set to $\lambda_0=1\,\rm \mu m$ for the numerical applications.
Our two beams cross at a total angle $2\theta=24^{\circ}$ and are initialized by the exact solution Eq.~\eqref{exact field} of the paraxial equation and with synchronized modulators, $t_0=0$. Both beams are linearly \(P\)-polarized, meaning that their electric fields lie in the \(x\text{--}y\) plane. The intersection of the two beams occurs at the center of the simulation box, with the average intensity at this point being \( I_i^0 = 40 \, \text{TW}/\text{cm}^{2} \). This corresponds to a normalized vector potential of \( a_i = 0.0134 \). The f-number is equal to 8, a focal length of \( f_0 = 8\,\mathrm{m} \) is employed and we consider $N=6$ RPP elements, corresponding to an FWHM of the intensity profile of the order of $48\,\lambda_0$. The temporal evolution of the laser beam fields is a ramp of width $2000\,\nu_0^{-1}$ followed by a plateau that lasts until the end of the simulation. The frequency modulators are here synchronized ($t_0=0$).\\
The parameters that we use ensure the reasonable numerical convergence of our PIC simulations, as shown in Ref.~\cite{OudinPOP2}. In subsections~\ref{V.3 vdy} and~\ref{V.4 w}, for SSD simulations, a modulation depth of \( M = 2 \) is used. All SSD simulations use a modulation frequency of $\nu_d = 14.25~\text{GHz}$, a cycle number of one, and a spectral cut-off of type $\Upsilon = \ceil{M}$. \\
In Figs.~\ref{I(x,y)}, we plot, at time $t = 56\,\text{ps}$, the intensity profile induced by the superposition of the two beams in a CH plasma. 
To estimate the exchanged power, we integrate the intensity profile of each beam along $(Oy)$ at the box input ($x \approx -400\,\mu$m) and at the box output ($x \approx 400\,\mu$m). We obtain power values at both the input and output of the box for each laser beam, denoted $P^0_1, P^0_2$ and $P^F_1, P^F_2$, respectively. The exchanged  power between the beams in the simulation is then calculated as $\frac{P^F_1 - P^F_2}{P^0_1 + P^0_2}$. After the growth of IAWs over a time $t \approx \nu^{-1}$, we obtain an exchanged power for spatially smoothed beams independent on time, as established in Ref.~\cite{OudinPOP2}. In contrast, for temporally smoothed beams, a significant temporal fluctuation of the exchanged power is observed, which was previously highlighted in Ref.~\cite{Huller}, and is confirmed as shown in Fig.~\ref{I(t)}. For simplicity, we will compare our linear theory in Eq.~\eqref{mean t} to the exchanged power values averaged over a modulation period from Smilei simulations.
The amplitude of these fluctuations as predicted by the simulations could be confronted to the model predictions  [Eq.~\eqref{fluctuation t}],  however, it does not add much to the discussion. In Figs.~\ref{I(t)}, we observe the temporal variation of the exchanged power, represented by the purple and green solid curves, respectively, for the flow-induced and frequency shift-induced power transfer. As for the RPP scenario, we observe that, for SSD beams, wavelength detuning leads to a reduction of the power exchanged between the beams near acoustic resonance.

\subsection{\label{V.2 vdx}
\DB{Crossing of two RPP beams with the same central frequency when there is a flow in the (xOy) plane}}
\begin{figure}
\centering
\caption*{\hspace{6 mm}$ v_{dy}\neq0 \quad v_{dz}=0 \quad \omega_1=\omega_2$ }\vspace{-1.0em}
\includegraphics[scale=0.32]{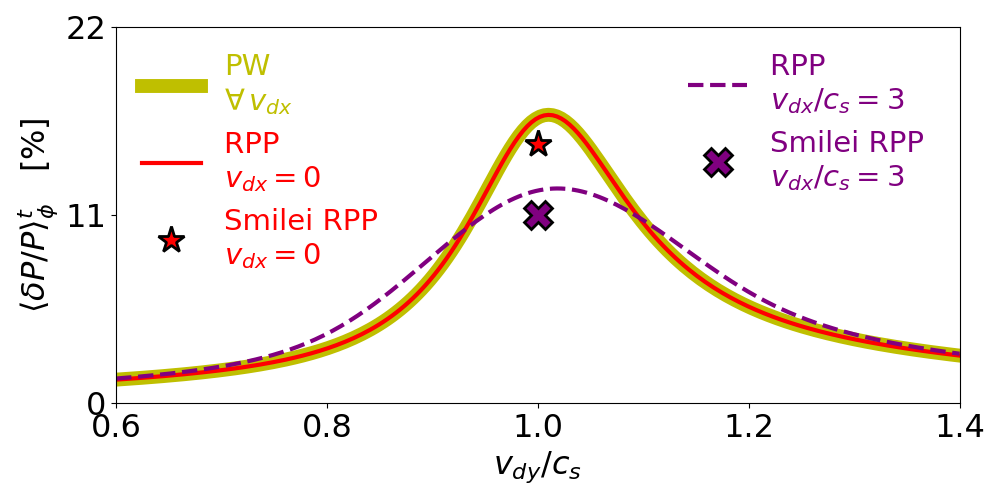}
\caption{\colorbox{Gray!20}{\parbox{\dimexpr\linewidth-2\fboxsep}{\textbf{FIG \thefigure:} Power exchanged between the beams in a CH plasma predicted by Eq. \eqref{mean t} (lines) and extracted from the steady-state regime of our PIC simulations (markers) is plotted against  \( v_{dy}/c_s \). The RPP case, where the plasma drifts at the sound speed $c_s$ in the main IAW direction (\(Oy\)), is shown in red. The RPP case, where the plasma drifts at \(3c_s\) in the direction orthogonal to the IAWs (\(Ox\)) and at the sound speed in the main IAW direction (\(Oy\)), is shown in purple. The plane wave case is represented by the yellow line.}}}
\label{vdx case validation}
\end{figure}
In the following simulations, \DB{where} the plasma drifts in the \( (xOy) \) plane, we employ reflective boundary conditions longitudinally and periodic boundary conditions transversely. 
Since we cannot apply periodic boundary conditions longitudinally to both the fields and the plasma simultaneously in the Smilei code (due to the beams' longitudinal propagation direction within the simulation box), we place two vacuum layers at the longitudinal edges of the simulation. Figure~\ref{vdx case validation} presents resonance curves for a CH plasma as predicted by the model given in Eq.~\eqref{mean t}, within the framework of spatially smoothed beam descriptions. 
The crossing of two plane waves (two RPP beams) in a plasma drifting along the ion acoustic wave IAW $y$-direction is shown as a yellow solid (dotted red) line. The crossing of two RPP beams in a plasma drifting at a velocity of \(c_s\) along \((Oy)\) and \(3c_s\) along \((Ox)\) is represented by the dashed purple curve\sout{s}. The red stars and purple crosses indicate the corresponding PIC simulation results, noting that the plane-wave model has been validated in Ref.~\cite{Debayle_op}. \\
For CH plasma, within the spatial smoothing framework, we observe a 20\%-reduction in exchanged power when the plasma drifts with a finite $v_{dx}$-component, compared to the RPP case where $v_{dx}=0$ (equivalent to the plane wave case without wavelength detuning). 
We observe, in this case, a quantitative agreement between the PIC simulations and the linear model. 

\subsection{\label{V.3 vdy}
\DB{Crossing of two RPP-SSD beams with the same central frequency when there is a drift in the O$y$ direction}}

\begin{figure*}
\centering
\begin{tabular}{ccc}
(a) $v_{dx}=v_{dz}=0 \quad v_{dy}\neq0  \quad \omega_1=\omega_2$ & (b) $v_{dx}=v_{dz}=v_{dy}=0 \quad \omega_1\neq\omega_2$ &  (c) C$^{6+}$-distribution function\\
\includegraphics[width = 0.29 \linewidth]{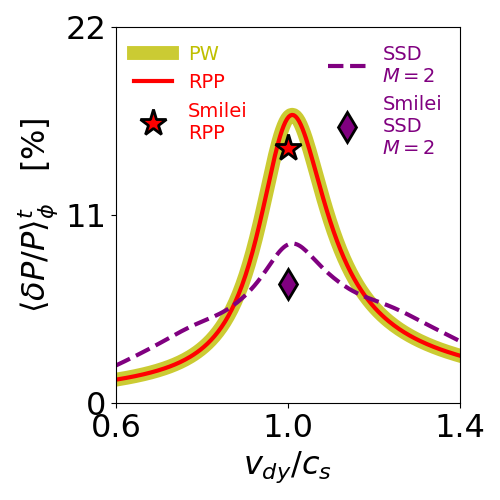}&
\includegraphics[width = 0.29 \linewidth]{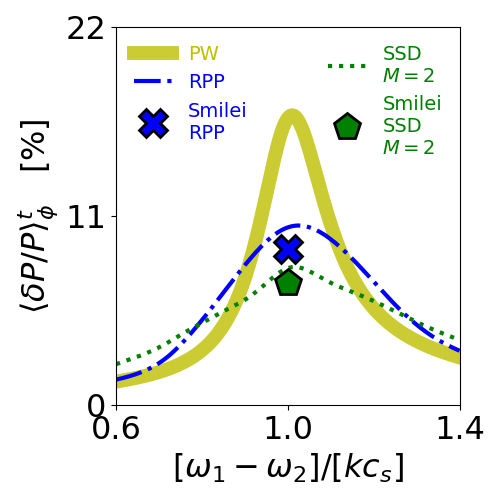}&
\includegraphics[width = 0.29 \linewidth]{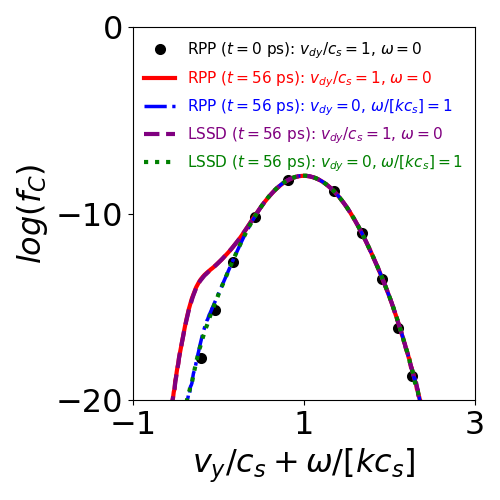} 
\end{tabular}
\vspace{-1.5em}
\caption{\colorbox{Gray!20}{\parbox{\dimexpr\linewidth-2\fboxsep}{\textbf{FIG \thefigure:} Power exchanged between the synchronized beams ($t_0=0$) in a CH plasma as predicted by Eq.~\eqref{mean t} (lines) and extracted from the steady-state regime of our PIC simulations (markers), shown as a function of (a) the plasma drift velocity along the acoustic-wave propagation direction, normalized to the ion-acoustic speed $v_{dy}/c_s$, or (b) the wavelength mismatch between the beams, normalized to the ion-acoustic frequency $(\omega_1-\omega_2)/(k c_s)$. In panel (a), the RPP and LSSD with \(M = 2\) cases are represented in red, and purple, respectively, while in panel (b), the same cases are shown in blue, and green, respectively. (a,b) The plane-wave case obtained from the model is represented by the yellow line. (c) Carbon-ion distribution function averaged over $v_x$, $y$ and $x$ in the vicinity of the crossing region. Black markers correspond to the initial distribution, while the other curves follow the same colour code as in (a,b).}}}
\label{Smilei comp}
\end{figure*}

This subsection presents simulations of CBET between two LSSD beams without wavelength detuning in drifting CH plasmas. In these simulations, the plasma drifts solely along the $(Oy)$ direction. Transverse periodic and longitudinal reflective boundary conditions are applied to the plasma, while longitudinal Silver--Müller and transverse periodic boundary conditions are imposed on the fields. 
Figure~\ref{Smilei comp}(a) presents the power exchange predicted by Eq.~\eqref{mean t} for synchronized beams ($t_0 = 0$), modelled as plane waves (thick yellow solid curve), RPP beams (red solid curves), and LSSD beams with $M=2$ (purple dashed curve), as a function of the plasma drift velocity along the main IAW propagation direction $(Oy)$ and  normalized to the ion acoustic speed $v_{dy}/c_s$. These theoretical predictions are compared with Smilei PIC simulations of the corresponding SSD configurations. We recall that the validation of the plane-wave case has been reported in Refs.~\cite{Debayle_op,OudinPRL,OudinPOP2}. The purple diamond and the red star represent simulation results for LSSD with \( M = 2 \), and for RPP beams, respectively.  
Figure~\ref{Smilei comp}(a) highlights that beams modelled as plane waves exchange the same power as spatially smoothed beams, whereas temporally smoothed beams transfer two to three times less energy near resonance. Close to the acoustic resonance ($v_{d||IAW}/c_s \approx 1$) and for the imposed modulation bandwidth, temporal smoothing significantly reduces the exchanged power compared with either spatially smoothed beams or plane waves. Away from resonance, however, temporal smoothing increases the exchanged power relative to the plane-wave and RPP cases.  \\
For a CH plasma at acoustic resonance [Fig.~\ref{Smilei comp}(a)], discrepancies of up to 10\% are observed between our linear predictions and the results from Smilei simulations, \DB{for} both RPP and SSD beam configurations. \DB{In these situations we note that as, illustrated in Figure~\ref{Smilei comp}(c), the carbon-ion distribution function significantly deviates from a Maxwellian, an effect we cannot account for in our theory.}

\subsection{\label{V.4 w} 
\DB{Crossing of two RPP-SSD beams with different central frequencies}}
This subsection presents simulations of CBET in CH plasmas at rest in the laboratory frame, using wavelength-shifted LSSD beams. Reflective boundary conditions are imposed on the plasma, while absorbing boundary conditions are applied to the electromagnetic fields.  
Figure~\ref{Smilei comp}(b) displays the power exchange predicted by Eq.~\eqref{mean t} for synchronized beams ($t_0 = 0$), modeled as plane waves (thick yellow solid curve), RPP beams (blue dash-dotted curve) and LSSD beams with $M=2$ (green dotted curve), plotted as a function of the wavelength shift between the beams, normalized to the ion acoustic frequency [$\omega/(kc_s) = (\omega_1 - \omega_2)/(kc_s)$].  
We recall that validation of plane wave configuration were previously reported in Refs.~\cite{Debayle_op,OudinPRL,OudinPOP2}.
Including spatial smoothing (blue dash-dotted curve) decreases $\delta P/P$ by  up to 40\% in CH plasmas compared to the plane wave model (yellow solid curve) near acoustic resonance. Away from resonance, the plane wave model tends to slightly underestimate the exchanged power.  
Introducing the SSD model further amplifies these differences: the RPP description then overestimates the exchanged power close to resonance and underestimates it far from it. The quantitative agreement between theoretical curves and PIC simulation points confirms the reduction of power exchange at resonance induced by realistic smoothing.  \\
\DB{Note that, as may be seen in Fig.~\ref{Smilei comp} (c), when there is a frequency mismatch between the beams, the ion velocity distribution function deviates much less from a Maxwellian than when the beams have the same central frequency. This comes with a better agreement between the theoretical and PIC simulation results. We also note that, whether there is a frequency mismatch or not, the deviation from a Maxwellian is not affected by SSD. An in-depth investigation of nonlinear kinetic effects and of their impact on the power exchange is out of scope of this paper, and is left for future work.}
\cite{Nguyen_2021, Seaton1,Yin_2023b}.
Nevertheless, our PIC simulations confirm that the linear model presented here remains accurate for density fluctuations of a few percent of the initial density (until power exchange levels of approximately $15$--$20\%$).

\section{\label{VI Conclusion et perspectives}Conclusion and perspectives} 
In this study, we have demonstrated  that spatial and temporal smoothing reduce the power exchanged between beams close to the acoustic resonance and increase it far from resonance, compared to a model considering the crossing of two plane waves. This result is supported by both Particle-in-cell and paraxial hydrodynamic simulations. We demonstrate that in the context of spatially smoothed beams, these deviations are attributed to the presence of a flow perpendicular to the main direction of propagation of the acoustic waves $(Ox)$ in 2D ($(Oz)$ in 3D) and to the Doppler shift between the beams. The deviations induced by plasma drift in the direction orthogonal to the beam crossing plane suggest the need to take into account a three-dimensional CBET model between RPP beams.
When the beams are smoothed both spatially and temporally, we observe important deviations when accounting for the SSD spatial dispersion component. We show that the wavelength shift and the flow orthogonal to the main IAW propagation direction have a subdominant effect compared to dispersion. Thus, in inline CBET models, describing SSD beams while accounting only for plasma flow along the IAW direction  is sufficient to estimate the exchanged power with good accuracy, provided that the SSD operates with a sufficiently large modulation depth and that the flow in directions orthogonal to the IAWs is not too strong. \\
The IAW frequency driven by the CBET  increases with the crossing angle and as a consequence, \DB{so does the acoustic damping rate}. This indicates that the deviation from the plane wave power exchange is less significant at large incident angle because all the beating wavelets may lie inside the resonance. The no-SSD case of Ref. \cite{OudinPOP2} shows that drift and frequency shift power exchange behave similarly at  large crossing angle. With SSD and arbitrary flow vectors, the CBET resonance width is an increasing function of the laser bandwidth and flow components, hence, the power exchange will deviate from the plane wave limit for large enough misaligned flows or laser bandwidth. A quantitative analysis of the plasma and laser parameters for which CBET with laser smoothing deviates from the plane wave models is given in Ref. \cite{Lettre_lalaire}.\\
We also evidenced, for the first time, an important sensitivity of the power exchange on the synchronisation of the laser beam SSD modulators. An averaging over the synchronization must be performed in the absence of exact knowledge of the time delays between each modulator. This effectively smooths out dispersion effects but can lead to significant prediction errors. In particular, for synchronized or desynchronized beams (with known time delays), the dispersion accounted for in our model must be included in the CBET inline model to capture the non-negligible deviations from a spatially smoothed beam model. 

We addressed the crossing of two beams; the extrapolation of our results to a multi-beam (cone) configuration is left for future work. Note that the inclusion of this model in the Troll code~\cite{Lefebvre,Debayle_2025_CBET} is underway, and a detailed study of its influence on implosion symmetry is left for a future study.
We also confirmed the importance of non-linear effects on the CBET as previously analyzed \cite{Nguyen, Seaton1, Yin_2023b}, however, our linear predictions seems to be satisfying for power exchange around $15-20\%$ and for the parameters addressed in this study. 

\appendix

\section{\label{app3 Derivation de l'échange}Derivation of the Exchanged Power Between SSD Beams in Steady State Regime}
\subsubsection{\label{II.4.1 Fluctuations temporelles}CBET time fluctuations in steady state}
The electron fluid conservation equation coupled to the total laser beam field propagation equation gives, in the framework of linear perturbation theory ($\textbf{A}_{i}=\textbf{A}_{i}^0+\delta \textbf{A}_{i}$), on long time scales, the laser beam field transport equation through the acoustic grating \cite{OudinPOP1,Seaton2,OudinPOP2}:
\begin{equation}
    i[\textbf{k}_1 \cdot \grad]\delta \textbf{A}_1=\frac{\omega_{pe}^2}{4N_{e0}c^2} \delta n_e \textbf{A}_2^0.
    \label{Eq transport}
\end{equation}
Introducing \(I_i = I_i^0 + \delta I_i\) as the perturbed, linearized beam intensity, where \(I_i^0 = A_{i,0}^2 \epsilon_0 \omega_0^2 v_g / 2\) is the initial intensity (assumed identical for the two beams), and \(\delta I_i = \mathrm{Re}\{A_i^0 \delta A_i^*\} \epsilon_0 \omega_0^2 v_g\) represents the linear fluctuation of intensity around equilibrium. Multiplying the above transport equation by $\textbf{A}_1^{0*}$, adding its conjugate equation and integrating using the characteristics method:
\begin{equation}
    (x,y,z) \rightarrow \textbf{r}'=(\mathcal{R}\cos(\theta),y_0+\mathcal{R}\sin(\theta),z_0) \ ,
\end{equation}
with \(\mathcal{R} \in [-R/2, +R/2]\), where \(R\) is the longitudinal extent of the beam crossing region. We compute the energy exchange between the upper and lower beams as a function of the constants \(y_0\) and \(z_0\), as detailed in~\cite{OudinPOP2}. These constants are then integrated over the width of the beam crossing area (BCA), noted $L$. Such that $y_0,z_0\in [-L/2,+L/2]$ gives the power exchange between the upper and lower beams integrated over the BCA:
\begin{align}
    &\frac{\delta P(t)}{P}=-\frac{I_2^0C}{N^{2\mathcal{D}}} \Im \sum_{\textbf{n}_{1},m_{1}}\sum_{\textbf{n}_{2},m_{2}}\sum_{\textbf{n}_{3},m_{3}}\sum_{\textbf{n}_{4},m_{4}} \prod_{j=1}^{4}
    \nonumber \\ & 
    \times J_{m_j}(M)e^{i\big\{ [\phi_{\textbf{n}_1}+\phi_{\textbf{n}_4}-\phi_{\textbf{n}_2}-\phi_{\textbf{n}_3}] -[\Omega_{m_{1}m_{2}}-\Omega_{m_{3}m_{4}}]t  \big\}}\nonumber \\& 
    \times f(\Omega_{m_1m_2},\textbf{K}_{\textbf{n}_{1}\textbf{n}_{2}})  e^{i\big\{m_4\psi^{\text{T/L}}_{\textbf{n}_4}-m_2\psi^{\text{T/L}}_{\textbf{n}_2}\big\}}  
    \nonumber \\&
    \times e^{i\big\{m_1\psi^{\text{T/L}}_{{\textbf{n}_1}}-m_3\psi^{\text{T/L}}_{{\textbf{n}_3}}-\omega_dt_0\big\}\big\{m_1-m_3\big\}} h_{\textbf{n}_1\textbf{n}_2\textbf{n}_3\textbf{n}_4} \ ,
    \label{moy y0}
\end{align} 
where $C=\frac{(\textbf{u}_1\cdot\textbf{u}_2)^2}{2\epsilon_0 \omega_0^3v_{g}^2} \, \frac{e^2}{m_e^2}$ , $P_i^0=I_i^0L$ is the initial power of beam i, $\delta P_i=\delta I_i^0L$ is the linear variation in power per unit length of beam i. We define $\delta P=\delta P_1-\delta P_2$ and $P=P_1^0+P_2^0$ respectively as the power difference and the total power of the two beams. We assume $\delta P_1=-\delta P_2$ which leads to variations in the power of the beams expressed as $\delta P/P=\delta P_1/P_1^0=-\delta P_2/P_2^0$. The term  $h_{\textbf{n}_1\textbf{n}_2\textbf{n}_3\textbf{n}_4}$ includes the three spatial quadratures and fulfills~\cite{OudinPRL}:
\begin{align}
    &h_{\textbf{n}_1\textbf{n}_2\textbf{n}_3\textbf{n}_4}=\frac{\int_{-\frac{L}{2}}^{\frac{L}{2}}\int_{-\frac{L}{2}}^{\frac{L}{2}}\int_{-\frac{R}{2}}^{\frac{R}{2}} e^{i[\textbf{K}_{\textbf{n}_1\textbf{n}_2}-\textbf{K}_{\textbf{n}_3\textbf{n}_4}].\textbf{r}'} dy_0dz_0d\mathcal{R}}{L^2}\nonumber \\&
    =\frac{L}{\sin(2\theta)}\sinc\Big\{\frac{L}{2}\big[k_{\textbf{n}_2}-k_{\textbf{n}_4} \big]\cdot\hat{\textbf{e}}_y \Big\} \nonumber \\&  
    \times \sinc\Big(\frac{L\cos(\theta)}{2}\big[k_{\textbf{n}_1}+k_{\textbf{n}_4}-k_{\textbf{n}_2}-k_{\textbf{n}_3} \big]\cdot\hat{\textbf{e}}_y\Big\}  \nonumber \\& 
    \times \sinc\Big\{\frac{L}{2}\big[k_{\textbf{n}_1}+k_{\textbf{n}_4}-k_{\textbf{n}_2}-k_{\textbf{n}_3} \big]\cdot\hat{\textbf{e}}_z\Big\}\ .
\end{align}
This power exchange depends both on the RPP elements sampling and (unlike the RPP case) on time. 
Thereafter, by addressing small crossing angles $\theta \approx 10^\circ$ , we will neglect the influence of the RPP elements sampling and study only the time influence on this exchanged power. In the experiments, the number of RPP elements being of the order of a million, the statistical variance effects of the model do not reflect experimental reality~\cite{OudinPOP2}. 
Since the elements of the phase plates are all decorrelated~\cite{RuyerSBS,RuyerSpray,charles_2025}:
\begin{equation}
    \langle e^{i[\phi_{\textbf{n}_1}+\phi_{\textbf{n}_4}-\phi_{\textbf{n}_2}-\phi_{\textbf{n}_3}]}\rangle_\phi=\delta_{\textbf{n}_1,\,\textbf{n}_3}\delta_{\textbf{n}_2,\,\textbf{n}_4}
\end{equation}
with $\langle \rangle_\phi$ denoting an average over an infinite number of RPP realizations, and $\delta_{i,\,j}$ the Kronecker symbol.
The physical image of this average is to considered only the couplings between ponderomotive and acoustic gratings that results from the same pair of phase plate elements. This amounts to taking $\textbf{n}_1=\textbf{n}_3$ and $\textbf{n}_2=\textbf{n}_4$ in the expression of the statistical variance on the exchanged power Eq.~\eqref{fluctuation t}, we obtain:

\begin{align}
    &\frac{\langle\delta P(t)\rangle_\phi}{P}=-\frac{P_2^0C}{\sin(2\theta)N^{2\mathcal{D}}}\sum_{\textbf{n}_{1},m_{1}}\sum_{\textbf{n}_{2},m_{2}}\sum_{m_{3}}\sum_{m_{4}} \prod_{j=1}^{4}
    \nonumber \\ &
    \times J_{m_j}(M) e^{-i\omega_dt\{m_1+m_4-m_2-m_3\}}f(\Omega_{m_1m_2},\textbf{K}_{\textbf{n}_{1}\textbf{n}_{2}})
    \nonumber \\&
    \times e^{i[\{\psi^{\text{T/L}}_{\textbf{n}_1}-\omega_dt_0\}\{m_1-m_3\}-\psi^{\text{T/L}}_{\textbf{n}_2}\{m_2-m_4\}]} \ .
    \label{fluctuation t}
\end{align} 
This power exchanged at steady state depends on time periodically, the period is the period of the frequency modulator $2\pi/\omega_d$.

\subsubsection{\label{II.4.2 Puissance moyenne}Average steady-state CBET value}
Let's now average the time-dependent power exchange over one laser beam modulation period $t_d=\nu_d^{-1}$ using :
\begin{equation}
    \langle e^{-i\omega_d\{m_1+m_4-m_2-m_3\}t}\rangle_{t_d}\approx\sinc(\pi(m_1+m_4-m_2-m_3)) \ .
\end{equation}
This sinc term is negligible if $m_1+m_4-m_2-m_3>\pi$, which is fairly well verified for a large number of colours. For SSD with $M<2$ the error induced by this assumption is negligible, of the order of a few percent. Thus, in the limit \(M \gg 1\),
\begin{equation}
    \sinc\big(\pi(m_1+m_4-m_2-m_3)\big)\approx\delta_{m_1+m_4,\, m_2+m_3} \ .
\end{equation}
This corresponds to setting \(m_4 = m_3 + m_2 - m_1\) in Eq.~\eqref{fluctuation t}, which yields Eq.~\ref{mean t} in Section~\ref{III.4 Puissance échangée}.
\begin{align}
    &\frac{\langle\delta P\rangle_\phi^t}{P}=-\frac{P_2^0C}{\sin(2\theta)N^{2\mathcal{D}}}\sum_{\textbf{n}_{1},m_{1}}\sum_{\textbf{n}_{2},m_{2}}\sum_{m_{3}} \prod_{j=1}^3 J_{m_j}(M)
    \nonumber \\ &
    \times J_{m_2+m_3-m_1}(M) f(\Omega_{m_1m_2},\textbf{K}_{\textbf{n}_{1}\textbf{n}_{2}})  
    \nonumber \\&
    \times e^{i\{\psi^{\text{T/L}}_{\textbf{n}_1}-\psi^{\text{T/L}}_{\textbf{n}_2}-\omega_dt_0\}\{m_1-m_3\}} \ .
    \label{mean t annexe}
\end{align}  
\section{\label{app2 Smilei model}Spatially and temporally smoothed laser beam fields implemented in the Smilei PIC code}
The field smoothed by longitudinal (respectively, transverse) spectral dispersion corresponds to the superposition of \(2M+1\) wavelet packets, each diffracted by the RPP and characterized by a different focal length $F_{m_{L}} = f_0 \left(1 + N_c m_{L} \nu_d/\nu_0\right)$ (respectively, each having a distinct transverse focal point $y^{\mathrm{foc}}_{m_{T}} = f_\# \lambda_0 m_{T}$).
Considering the coordinate system \((Oxyz)\), with its origin placed at a distance \(f_0\) from the RPP (composed of \( N^2 \) phase plate elements), as introduced in Sec.~\ref{III.1 Champs rpp ssd}, the field resulting from the resolution of the Fresnel integral can be expressed at any point in space as:
\begin{align}
    \tilde{\textbf{A}}_{\text{F}/\pm}&=A_\text{L}\textbf{u} \sum_{m_T=-M}^{M}\sum_{m_L=-M}^{M}J_{m_L}(M)J_{m_T}(M)
    \nonumber \\& \times e^{i\big\{m_T+m_L\big\}\big\{\frac{\bar{x}}{c}-t\big\}\omega_d}A_{\text{F}/\pm}^{\text{RPP}}(m_T,m_L) \, .
    \label{exact field}
\end{align}
We define \(\bar{x} = x + f_0\) to simplify the expressions. The superscripts \(+\) and \(-\) denote the field evaluated for \(x > 0\) and \(x < 0\), respectively, corresponding to positions after and before the focal point of the central spectral component, located at a longitudinal distance \(f_0\) (along the laser beam propagation) from the RPP. The subscript \(F\) indicates evaluation exactly at \(x = 0\), i.e., at the focal point. The RPP field of each wavelet packet, perturbed by phase dispersion and modulation, can be expressed at any point with \(x \neq 0\) as:
\begin{align}
    \tilde{A}&^{\text{RPP}}_{\pm}(m_{T},m_{L})=\Big[\frac{e^{-i\frac{\pi(1\pm1)}{4}}}{2}\sqrt{\frac{F_{m_{L}}}{|x-f_{m_{L}}|}}\Big]^{\mathcal{D}} \nonumber \\
    & \times e^{i\frac{k_0\big\{y-y_{m_{T}}^{\text{foc}}\big\}^2}{2(x-f_{m_{L}})}} \sum_{n_{y}=1}^N\sum_{n_{z}=1}^N e^{i\phi_{n_{y}n_{z}}}\chi_{n_{y}m_{L}m_{T}}^{\pm}\kappa_{n_{z}m_{L}}^{\pm} \, .
    \label{en tout point sauf foc}
\end{align}
Where \(f_{m_{L}} = m_{L} f_0 \nu_d / \nu_0 = F_{m_{L}} - f_0\) represents the focal length perturbation associated with each wave packet of the spectrum modulated by the focusing grating.  
At the focal point \(x = 0\), the exact SSD field is given by:
\begin{align}
    \tilde{A}&^{\text{RPP}}_{F}(m_{T},m_{L})=\Big[\frac{e^{-i\frac{\pi}{4}}}{\sqrt{2\pi}}\sqrt{\frac{k_0d^2}{F_{m_{L}}}}\Big]^\mathcal{D} e^{i\frac{k_0\big\{y^2-y_{m_{T}}^{\text{foc}^2}+z^2\big\}}{2F_{m_{L}}}}\nonumber\\&\times \sinc\Big\{\frac{k_0d(y-y_{m_{T}}^{\text{foc}})}{2F_{m_{L}}}\Big\}\sinc\Big\{\frac{k_0dz}{2F_{m_{L}}}\Big\}\nonumber\\&\times\sum_{n_{y}=1}^N\sum_{n_{z
    }=1}^N e^{i\big\{\phi_{n_{y}n_{z}}-\frac{k_0d}{F_{m_{L}}}\big[(y-y_{m_{T}}^{\text{foc}})n_{y}+zn_{z}\big]\big\}} \, .
    \label{à la foc}
\end{align}
Where we have defined:
\begin{align}
    &\chi_{n_{y},m_{L},m_{T}}^\pm=\erf\Big\{e^{\pm i\frac{\pi}{4}}K_{m_{L}}\big[a_{n_{y}+1}-\frac{F_{m_{L}}}{x-f_{m_{L}}}(y\nonumber\\&-\frac{\bar{x}}{F_{m_{L}}}y_{m_{T}}^{\text{foc}})\big]\Big\}-\erf\Big\{e^{\pm i\frac{\pi}{4}}K_{m_{L}}\big[a_{n_{y}}-\frac{F_{m_{L}}}{x-f_{m_{L}}}(y\nonumber\\&-\frac{\bar{x}}{F_{m_{L}}}y_{m_{T}}^{\text{foc}})\big]\Big\}\, , \\
    & a_{n_{y}}/d=n_{y}-N/2-1 \, ,\\
    &\kappa_{n_{z},m_{L}}^\pm=\erf\Big\{e^{\pm i\frac{\pi}{4}}K_{m_{L}}\big[b_{n_{z}+1}-\frac{F_{m_{L}}}{x-f_{m_{L}}}z\big]\Big\}\nonumber\\&\quad-\erf\Big\{e^{\pm i\frac{\pi}{4}}K_{m_{L}}\big[b_{n_{z}}-\frac{F_{m_{L}}}{x-f_{m_{L}}}z\big]\Big\}\, , \\
    & b_{n_{z}}/d=n_{z}-N/2-1\, ,\\
    & K_{m_{L}}=\sqrt{k_0\frac{|x-f_{m_{L}}|}{2\bar{x}F_{m_{L}}}}\, .
\end{align}
This exact field is implemented in Smilei PIC code.  
For transverse SSD smoothing, we recover Eq.~\eqref{champ ref rpp} by considering the field exactly at \(x=0\) as given in Eq.~\eqref{à la foc}, whereas for longitudinal SSD smoothing, we perform an expansion in the limit \(x \xrightarrow[\neq]{} 0\) of Eq.~\eqref{en tout point sauf foc}. The main assumptions leading to the simplified expression Eq.~\eqref{champ ref rpp} are detailed in Refs.~\cite{OudinPOP1,OudinPOP2}.

\section*{Acknowledgements}
This work has been done under the auspices of  CEA-DAM and
the simulations were performed using HPC resources at TGCC/CCRT and CEA-DAM/TERA.

\section*{Data availability}
The data that supports the findings of this study are available from the corresponding authors upon reasonable request.

\bibliography{apssamp}
\end{document}